\documentclass[%
 reprint,
superscriptaddress,
longbibliography,
showkeys,
 amsmath,amssymb,
 aps,
pre, 
]{revtex4-2}

\usepackage{graphicx}
\usepackage[caption=false]{subfig} 

\usepackage{dcolumn}
\usepackage{bm}

\usepackage{hyperref}
\hypersetup{
    colorlinks=true,
    linkcolor=blue,
    filecolor=magenta,      
    urlcolor=cyan,
    citecolor = purple,
    pdftitle={Overleaf Example},
    pdfpagemode=FullScreen,
    }
\usepackage{array}
\usepackage{float}

\newcommand{\ee}{\end{equation}} 
\newcommand{\be}{\begin{equation}}

\usepackage[mathscr]{euscript}  
\usepackage{xcolor}  

\usepackage{tcolorbox}
\usepackage{comment}
\usepackage{float}

\makeatletter
\newsavebox{\@brx}
\newcommand{\llangle}[1][]{\savebox{\@brx}{\(\m@th{#1\langle}\)}%
  \mathopen{\copy\@brx\kern-0.5\wd\@brx\usebox{\@brx}}}
\newcommand{\rrangle}[1][]{\savebox{\@brx}{\(\m@th{#1\rangle}\)}%
  \mathclose{\copy\@brx\kern-0.5\wd\@brx\usebox{\@brx}}}
\makeatother

\newcommand{\rmd}{{\rm d}}

\begin{document} 

\preprint{APS/123-QED}

\title{Thermodynamic cost of finite-time stochastic resetting}

\author{Kristian St\o{}levik Olsen}\thanks{kristian.olsen@hhu.de}\thanks{Equal contributions}
\affiliation{Institut für Theoretische Physik II - Weiche Materie, Heinrich-Heine-Universität Düsseldorf, D-40225 Düsseldorf, Germany}

\author{Deepak Gupta}\thanks{phydeepak.gupta@gmail.com}\thanks{Equal contributions}
\affiliation{Department of Physics, Indian Institute of Technology Indore, 453552,Khandwa Road, Simrol, Indore, India}
\affiliation{Nordita, Royal Institute of Technology and Stockholm University, Hannes Alfvéns väg 12, 106 91, Stockholm, Sweden}

\author{Francesco Mori}
\affiliation{Rudolf Peierls Centre for Theoretical Physics, University of Oxford, Oxford, United Kingdom}

\author{Supriya Krishnamurthy}\thanks{supriya@fysik.su.se}
\affiliation{Department of Physics, Stockholm University, 106 91, Stockholm, Sweden}


\begin{abstract}
Recent experiments have implemented resetting by means of an external trap, whereby a system relaxes to the minimum of the trap and is reset in a finite time. In this work, we set up and analyze the thermodynamics of such a protocol. We present a general framework, even valid for non-Poissonian resetting, that captures the thermodynamic work required to maintain a resetting process up to a given observation time, and exactly calculate the moment generating function of this work. 
Our framework is valid for a wide range of systems, the only assumption being relaxation to equilibrium in the resetting trap. Examples and extensions are considered.
In the case of Brownian motion, we investigate optimal resetting schemes that minimize work and its fluctuations, the mean work for arbitrary switching protocols and comparisons to previously studied resetting schemes. Numerical simulations are performed to validate our findings.
\end{abstract}

\pacs{Valid PACS appear here} 
\maketitle

\section{Introduction}
\label{sec:intro}

Systems undergoing random resetting have been extensively studied over the last decade. Despite the simplicity of the concept, whereby a system is intermittently returned to its initial state, stochastic resetting continues to reveal intriguing phenomena more than a decade after its conception \cite{evans2011diffusion, evans2011optimal}.

The rich phenomenology of resetting has made it an attractive arena to explore the intricacies of non-equilibrium systems. In particular, steady states far from equilibrium have been studied in a wide range of systems, such as for resetting in potential landscapes \cite{Pal_PRE, ray2020diffusion,KSO2023}, in underdamped Brownian motion \cite{Gupta_2019}, in a random accelerated system~\cite{singh2020random}, and resetting mediated by fluctuating potentials \cite{gupta2020stochastic,mercado2020intermittent,santra2021brownian,mercado2022reducing, Gupta_2021_SR}. Steady states under variations of the resetting scheme have also been considered, such as time-dependent resetting rates \cite{pal2016diffusion,shkilev2017continuous}, non-instantaneous resets \cite{Evans_2019, Pal_19, radice2022diffusion, tucci2022first,19MPuigdellosas,20Brodova,giorgini2023thermodynamic}, and non-Poissonian waiting times \cite{eule2016non, nagar2016diffusion,radice2022diffusion}. Other fruitful avenues of research include investigating the way resets expedite search processes \cite{evans2011diffusion, bressloff2020search, reuveni2016optimal,pal2017first,chechkin2018random, pal2019first, ahmad2019first,ahmad2022first,tucci2022first,singh2022first,de2020optimization,besga2020optimal}, applications of large deviation theory \cite{monthus2021large, meylahn2015large, coghi2020large}, and the non-trivial relaxation dynamics towards steady states \cite{Gupta_2019, Sanjib_relax, singh2020resetting}. This is only a taste of the broad phenomenology of stochastic resetting that has been uncovered over the past years, and for a more comprehensive review see Ref.~\cite{evans2020stochastic}.

An essential, and often overlooked, aspect of systems that undergo resets is the inherent cost associated with performing the resetting and hence maintaining the non-equilibrium steady state. An understanding of the relationship between stochastic resetting and its associated thermodynamic cost is both necessary and timely, particularly in the light of various experimentally realized resetting protocols in recent years \cite{goerlich2023experimental,tal2020experimental,besga2020optimal}.

A natural framework for assessing the thermodynamic cost of resetting processes is {\it stochastic thermodynamics}, which provides a basis for defining and studying thermodynamic quantities and their fluctuations on the microscopic scale. This framework has had great success in analyzing both theoretical and experimental scenarios over the past couple of decades \cite{S12,ciliberto2017experiments}. The examination of resetting through the lens of stochastic thermodynamics, however, is a relatively recent endeavor. Such studies were first addressed in Refs.~\cite{fuchs2016stochastic, Entropy_UD}, where an average entropic contribution of resetting was identified. Stochastic resetting has further been shown to satisfy integral fluctuation theorems \cite{pal2017integral}, and work fluctuations have been investigated for a system with simultaneous particle and protocol resets \cite{gupta2020work}.  
Stochastic resetting has been found to be an efficient protocol affecting the onset of the Mpemba effect~\cite{Busiello_2021} and has also been used to derive thermodynamic bounds on the speed limit to connect two distributions ~\cite{SL-UD}. In the last couple of years, more studies on this topic have emerged, including thermodynamic work associated with resetting using potentials \cite{tal2020experimental, Deepak2022_work}, the  full distribution of entropy production when resetting positions are drawn from a distribution \cite{mori2023entropy}, trade-off relations between work and time \cite{Pal_tradeoff} and generic costs that may not be of a thermodynamic nature  but are determined by the distance traversed since the last reset \cite{Sunil_2023}. The thermodynamics of resetting also closely resembles that of particles in fluctuating potentials, as studied in \cite{alston2022non}.

In all of the above, thermodynamic cost of resetting is estimated either as total work done to implement a physical mechanism mediating the resetting for a trajectory of temporal length $t$ \cite{Deepak2022_work}, or as total heat dissipated (equivalently entropy produced) \textit{during} the resetting process over time $t$ \cite{mori2023entropy}. 
By the first law of thermodynamics, the averages of these two quantities are related, and we expect the first moments of heat and work to be the same, barring an internal energy contribution. In either manner of calculating the cost, the {\textit{time}} taken during the actual resetting is also clearly relevant. The cost of finite-time resetting has been studied for  a process where the resetting happens deterministically at finite speed or finite time \cite{tal2020experimental}. Another well-studied version of  finite-time resetting~\cite{gupta2020stochastic, Gupta_2021_SR} for which the cost has been computed is when resetting occurs via a first-passage process~\cite{Deepak2022_work,Pal_tradeoff}, {\color{black}and we stress that such a resetting protocol (in contrast to what we discuss in this paper) necessitates feedback control and is more challenging to
implement experimentally.}
{Experimentally, it is very natural
to implement finite-time resetting by means of a resetting trap generated, for example, by optical-tweezers methods~\cite{besga2020optimal,goerlich2023experimental}. To implement resets, the resetting trap is activated recurrently and each time it is activated, it is left to act for a duration which is long enough for the particle to thermalize~\cite{besga2020optimal,goerlich2023experimental}. Once thermalization occurs and the particle's
distribution has relaxed to the thermal Boltzmann state (characterized by the resetting potential), the particle has been reset. Note that in this case the particle is not reset to an exact location as is usually considered, but instead to a position drawn from the equilibrium
thermal distribution in the trap.}
It is interesting here too to understand the distribution of work costs for such a process, but to our knowledge the thermodynamic cost of resetting and its fluctuations has not been studied in this case.

In this paper, we study exactly this problem, for which we present a general framework for calculating the thermodynamic cost and its fluctuations for resetting processes in finite time. The framework we develop
holds for any underlying dynamics between resets, any spatial dimension, any resetting trap that is confining, and arbitrary durations of time intervals between resets.
It is in principle generalizable even to the case when the resetting trap takes a finite time to switch on, such as in a recent experimental study \cite{goerlich2023experimental}, though we leave that to a future study.

This paper is organized as follows. Section~\ref{sec:theory} presents the stochastic resetting scheme considered in this paper, and introduces the general stochastic thermodynamics framework which will be used to study its cost. Section~\ref{sec:fluctuations} derives an exact expression for the moment generating function of the work associated with resetting, and provides one of the main results of the paper. A recursive relation for the moments is derived, allowing a full characterization of the fluctuations of the thermodynamic cost. Section~\ref{opt-trap} considers a concrete example, namely resetting of a Brownian particle in two intermittent harmonic potentials alternating in time. The mean and variance of the work are calculated, and optimal resetting protocols are considered. Section~\ref{sec:finite} extends the analysis of the mean work to the case where the time duration of the sojourn, in between potential switches, is arbitrary. Hence, in section~\ref{sec:finite}, equilibration is \emph {not} assumed in the resetting trap. This allows us to make connections between resetting protocols, and intermittent fluctuating potentials more broadly. Section~\ref{sec:comp} provides a comparison between the proposed resetting scheme and other implementations of resetting studied in the past, before Sec.~\ref{sec:concl} provides a concluding discussion.

\begin{figure*}[t]
    \centering
    \includegraphics[width = 0.86\textwidth]{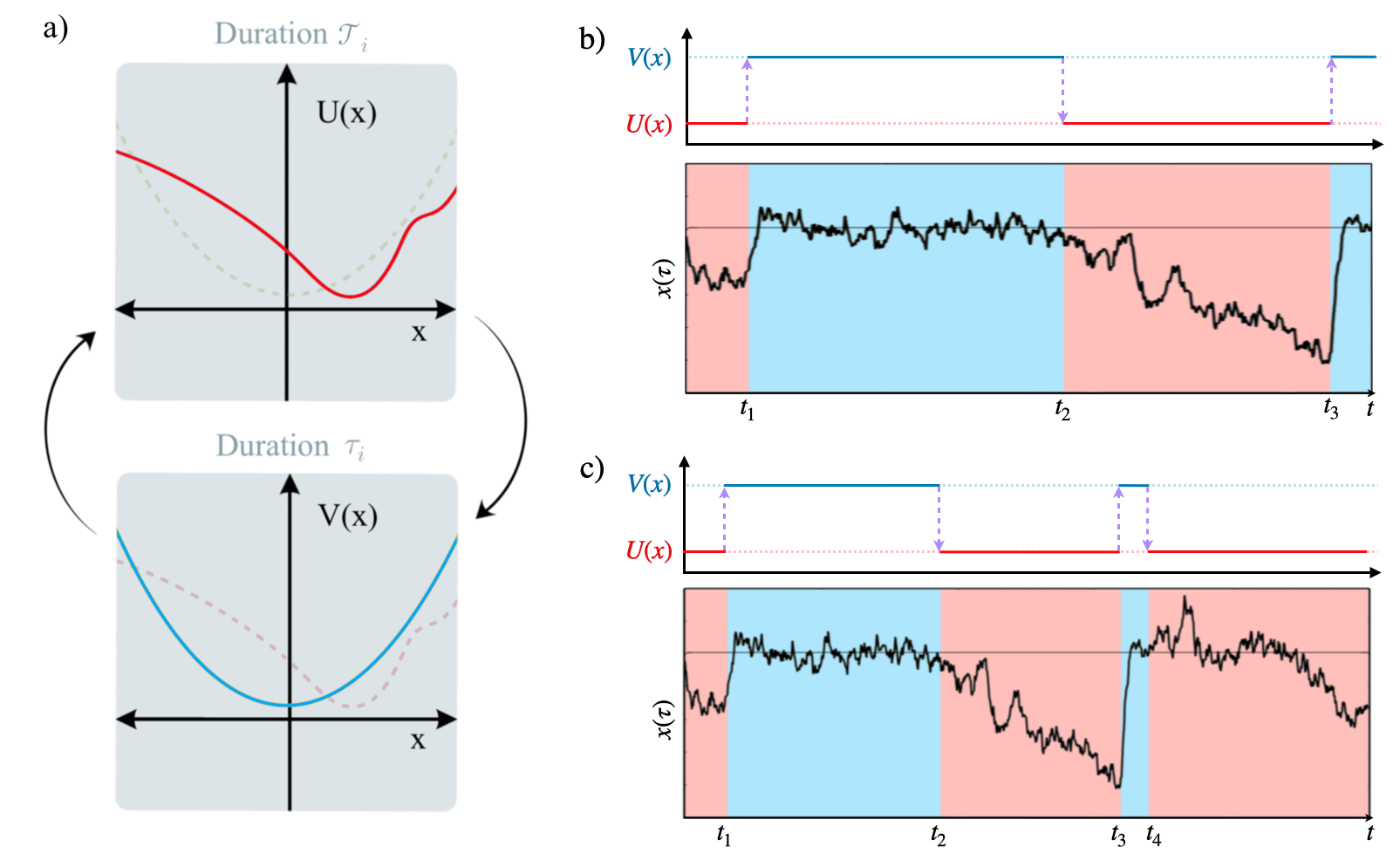}
    \caption{a) We consider a resetting protocol where the system intermittently switches between a resetting potential $V(x)$ and an exploration potential $U(x)$, both of which are completely general. b-c) Schematic of realizations of the resetting protocol as a function of time. The system is initially  prepared in the potential $V(x)$ (such that $p_{\rm eq}(x)\propto e^{-\beta V(x)}$), and switches to $U(x)$ at time $t=0$. At a fixed observation time $t$, the system may reside in either the resetting phase (panel b), or in the exploration phase (panel c).
    }
    \label{fig:scheme}
\end{figure*}

\section{General theory}
\label{sec:theory}
The simplest implementation of resetting consists of using a sharp time-independent confining potential $V(x)$ to move the particle to a desired resetting location, e.g., by having the (global) minimum of $V(x)$ at the desired reset position~\cite{besga2020optimal,goerlich2023experimental, mercado2020intermittent} {\it(return phase)}. In the interval between two resetting phases, the  particle diffuses under the influence of some other external potential $U(x)$  {\it(exploration phase)}. A schematic for this process is shown in Figs.~\ref{fig:scheme}a. 
In order to conform to recent experimental work on resetting  \cite{goerlich2023experimental}, we will refer to the potential $U(x)$ as the shallow potential, and $V(x)$ as the sharp potential or, equivalently, the resetting potential. However, most results in this paper are valid for arbitrary intermittent potentials, i.e., these do not necessarily need to be sharp and shallow. However, the resetting potential does need to be a confining potential.

Implementing such a protocol of resetting clearly has a  significant cost in terms of  thermodynamic work performed. 
In addition, as for all small-scale systems, the work performed will be dependent on the specific stochastic trajectory
followed and hence  all  moments of the work need to be estimated by averaging over an  ensemble of such trajectories.

To this end, we consider a single particle moving in $d$ dimensions under the influence of an external time-dependent potential $V_\text{tot}[x_\tau,\Lambda(\tau)]$, where $\Lambda(\tau)$ is a time-dependent control parameter. The dynamics reads
\begin{equation}
    \frac{dx_\tau}{d\tau}=-\gamma^{-1} \nabla V_\text{tot}[x_\tau,\Lambda(\tau)]+\eta(\tau)\,,
\end{equation}
where $\gamma$ is the friction constant. For now, $\eta(t)$ can be any additional forces acting on the particle that do not contribute directly to the work. For example, for the simplest case of a Brownian particle $\eta(\tau)$ are thermal forces represented as Gaussian white noise with zero mean and correlation function $\langle\eta_i(\tau)\eta_j(\tau')\rangle=2D\delta_{ij}\delta(\tau-\tau')$, with $i,j$ being spatial indices. Figures~\ref{fig:scheme}b-c display examples of an overdamped Brownian particle's trajectory under the above protocol.   We choose the potential $V_\text{tot}[x_\tau,\Lambda(\tau)]$ to be a convex linear combination of the resetting $V(x)$ and exploration $U(x)$ potentials:
\begin{equation}\label{concomb}
    V_\text{tot}[x_\tau,\Lambda(\tau)] \equiv \Lambda(\tau)~U(x_\tau) + [1-\Lambda(\tau)]~V(x_\tau)\ .
\end{equation} 
The dichotomous control parameter in Eq.~\eqref{concomb} follows, for $j\geq 1$,
\begin{equation}\label{lambda-tau}
    \Lambda(\tau) = 
\left\{
	\begin{array}{ll}
		1  & \mbox{if } \tau \in (t^+_{2j-2}\ ,~t^-_{2j-1})\ , \\
		0  & \mbox{if } \tau \in (t^+_{2j-1}\ ,~t^-_{2j})\ .
	\end{array}
\right.
\end{equation}
The time instance with a superscript $`-'$ ($`+'$) denotes a time just before (after) the switching of a potential (see vertical dashed arrows in the schematic~\ref{fig:scheme}b-c). Notice that these time instances depend upon how we externally control the system's dynamics by switching these potentials. We draw the time intervals $\mathcal{T}_{j} \equiv t_{2j-1}^- - t^+_{2j-2}$,
corresponding to the exploration phase, from a probability density function $f(\mathcal{T}_j)$. Similarly, the duration of the resetting phases is $\tau_j \equiv  t_{2j}^- - t_{2j-1}^+$ and is drawn from a density $h(\tau_j)$. See Fig.~\ref{fig:scheme}b-c) for a schematic.

The work done to reset the system is the net energy change at the instances
when one potential is instantaneously switched to another via Eqs.~\eqref{concomb} and~\eqref{lambda-tau}, i.e., during $U(x) \leftrightarrow V(x)$  (see vertical arrows in Fig.~\ref{fig:scheme}b-c). 
Then, summing over all switching instances, the total work in a single stochastic trajectory $\{x_\tau\}_0^t$ is~\cite{JE-PRL,Crooks1998}
\begin{align}
    W[\{x_\tau\}_0^t]  &\equiv \sum_{j=1} [U(x_{2j-2}^+) - V(x_{2j-2}^-) ]\  \nonumber\\
    &+ \sum_{j=1} [V(x_{2j-1}^+) - U(x_{2j-1}^-) ]\ , \label{eq:work}
\end{align} 
where for brevity we write $x_{k}^\pm \equiv x(t_k^\pm)$.
Notice that, in our convention, $W>0$ implies the work is performed on the system. Note also that this sum includes a work at time $t=0$. We have assumed that the system is initially prepared in a steady state associated with the resetting potential $V(x)$, which is switched off at time $t=0$.

During the implementation of the resetting protocol, the system exchanges heat with the environment. This heat exchange is the net energy change due to changes in the state of the system for a fixed control parameter, i.e., for a fixed $\Lambda$. Then, for a single stochastic trajectory, the heat exchange is   

\begin{align}\label{eq:diss}
     Q[\{x_\tau\}_0^t] &\equiv \sum_{j=1} [U(x_{2j-2}^+) - U(x_{2j-1}^-)] \nonumber\\
     &+ \sum_{j=1} [V(x_{2j-1}^+) - V(x_{2j}^-)] \ \nonumber\\
      &+~\begin{cases}
    V(x_N^+) - V(x(t)) &{\rm for}~x(t)\in {\rm RP}\ ,\\
    U(x_N^+) - U(x(t)) &{\rm for}~x(t)\in {\rm EP}\ ,
    \end{cases}
\end{align}
where $Q>0$ implies the heat is dissipated in the environment, and EP and RP respectively denote exploration- and return-phase. The last term on the right-hand side~\eqref{eq:diss} depends on the phase in which the system's trajectory finishes for each realization (see Fig.~\ref{fig:scheme}b and c). Here $x_N^+$ and $x(t)$, respectively, correspond to the initial and final states of the system in the final phase. For instance, in Fig.~\ref{fig:scheme}b, $N = 3$ 
when the system is in $V(x)$ (the return phase) for the final time,  and in Fig.~\ref{fig:scheme}c,
$N=4$ when  the system is in $U(x)$ (the exploration phase) for the final time.

Subtracting the heat~\eqref{eq:diss} from the work done~\eqref{eq:work} for each trajectory gives 
\begin{equation}
    \Delta E[\{x_\tau\}_0^t] = W[\{x_\tau\}_0^t] - Q[\{x_\tau\}_0^t]
\end{equation}
as the net change in the internal energy between the initial and final states of the system at the level of a stochastic trajectory {\it(the first law of thermodynamics)}, where
\begin{align}
    \Delta E[\{x_\tau\}_0^t]  & = -V(x_0^-) \\
    & + \begin{cases}
    V(x(t)) &{\rm for}~x(t)\in {\rm RP}\ ,\\
    U(x(t)) &{\rm for}~x(t)\in {\rm EP}\ ,
    \end{cases}\nonumber
\end{align}
as expected. Here, the first term $V(x_0^-)$  on the right-hand side is the initial potential energy. 

Note that since on average $\Delta E[\{x_\tau\}_0^t]$ is expected to be bounded, the average rate of work will be equal to the average rate of heat dissipated by the system in the long-time limit. However, we do not expect that the higher cumulants or the full distribution behave analogously, i.e., the fluctuations of heat dissipation and work could be different even though their asymptotic mean rates are equal~\cite{ZonCilibertoCohen}.

\begin{figure}[t!]
    \includegraphics[width = 8.6cm]{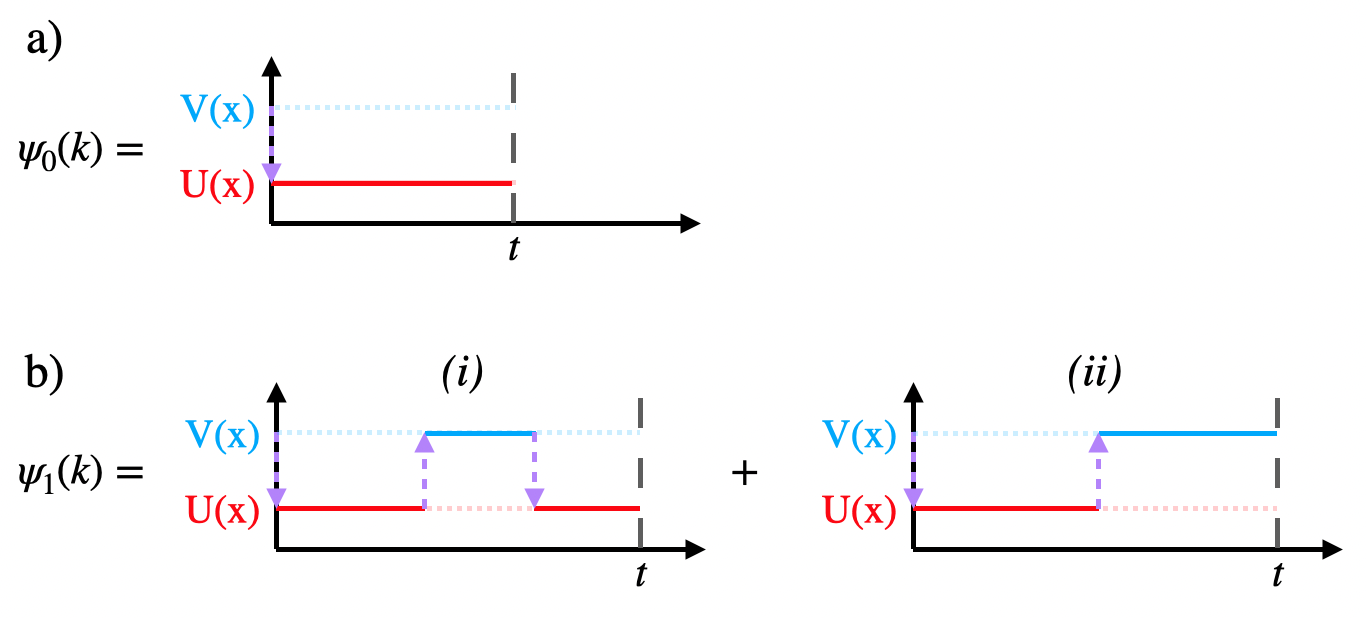}
    \caption{Contributions to the work generating function with zero $\psi_0(k)$ resets [panel a)] and one resetting $\psi_1(k)$ [panel b)].  $\psi_1(k)$ has two contributions, labelled $(i)$ and $(ii)$. }
    \label{fig:split}
\end{figure}

\section{Work fluctuations} 
\label{sec:fluctuations}

In this section, we investigate the  moment generating function of the work. To make analytical progress, we make the assumption that the 
particle equilibrates at the end of each resetting phase, i.e., that the durations $\tau_i$ (see Figs.~\ref{fig:scheme}b-c) are sufficiently long compared to the trap-relaxation time of the resetting potential $V(x)$. 
This is most easily ensured by choosing a deterministic duration for this return phase, e.g., $h(\tau) = \delta(\tau-\tau_\text{on})$, where we
parameterize the on-time as a multiple, $\theta$, of the relaxation time $\tau_\text{on} \equiv \theta \tau_\text{rel}$. This is an experimentally motivated scheme \cite{besga2020optimal,goerlich2023experimental}, which also has the theoretical advantage that for large values of the parameter ($\theta\gtrsim 1$), a renewal structure  emerges, with the particle
resetting to the Boltzmann state  $p_{\rm eq}(x)\propto e^{-\beta V(x)}$, after every reset.  While the results of this section are valid for any $h(\tau)$, they do rely on this renewal structure, which assumes that $h(\tau)$ is chosen accordingly.

In the following Sec.~\ref{sec:gen-func}, we derive the generating function of work fluctuations for a system undergoing the above stochastic resetting protocol for a fixed time $t$ (irrespective of the total number of resetting events occurring in that interval).
When the durations of both resetting and exploration phases are arbitrary, we cannot get the full generating function as the lack of renewal structure introduces arduous correlations, but we can nevertheless derive
an exact expression for 
the mean work per reset. This is done later in Sec.~\ref{sec:finite}.

\subsection{Generating function}
\label{sec:gen-func}

We develop a mathematical framework to investigate the stochastic work performed on the system during the resetting protocol. 
It is convenient to compute the moment generating function (characteristic function) of the work by considering the contributions arising from stochastic trajectories that involve different number of resets up to time $t$. Concretely, the moment generating function is written as
\begin{equation}
    C(k,t) \equiv   \left\langle  e^{ k W}\right\rangle_t\ = \sum_{n=0}^\infty~\int_{-\infty}^{+\infty}~\rmd W~\mathcal{P}_t(W,n)~e^{kW}\ , \label{mgf-def}
\end{equation}
where the average is calculated for a finite time $t$ over the entire set of stochastic trajectories undergoing the stochastic resetting protocol. In Eq.~\eqref{mgf-def}, $k$ is a Fourier variable conjugate to the work $W$, and $n$ indicates the number of resetting events in a given time $t${\color{black}, i.e., number of times the potential $V(x)$ is switched on.} {Indeed, in our model, a resetting event corresponds to the time interval in which the potential $V(x)$ is turned on (see Fig.~\ref{fig:split}).} The distribution $\mathcal{P}_t(W,n)$ is the joint distribution of work  $W$ and having $n$ resets in time $t$.  With this decomposition, we may write Eq.~\eqref{mgf-def} more compactly as
\begin{equation}\label{eq:genfunc_expression}
    C(k,t) = \sum_{n=0}^{\infty}~\psi_n(k,t)\ .
\end{equation}
In the following, we show the recipe to compute each of the contributions $\psi_n(k,t)$.
    \begin{itemize}
    \item The $n=0$ term in Eq.~\eqref{mgf-def} corresponds to no resetting event occurring over the entire observation time $t$ {\color{black}[see Fig. \ref{fig:split} a]}. The contribution to the moment generating function from this term is
    \begin{align}
    \psi_0(k, t) \equiv g(t) \int^{+\infty}_{-\infty}~\rmd x_0~p_{\rm eq}(x_0)~e^{k [U(x_0)-V(x_0)]}\ , \label{psi0} 
    \end{align}
    where $g(t) =  1-\int_0^t~dt'~f(t')$ is the probability that the system experiences no resetting up to time $t$. Initially, the system is prepared in an equilibrium state $p_\text{eq}(x_0)\propto e^{-\beta V(x_0)}$ with respect to the potential $V(x_0)$, so that at time $t=0$ the work done in switching from $V(x) $ to $U(x)$ is $U(x_0)-V(x_0)$.

    \item The $n=1$ term in Eq.~\eqref{mgf-def} corresponds to
    the case of one resetting event up to time $t$, i.e, switching on the sharp potential $V(x)$ once in that time-interval $t$. This term leads to  two contributions [see Fig. \ref{fig:split}b].
    \begin{itemize}
        \item[i)] The system starts from a Gibbs state $p_{\rm eq}(x)\propto e^{-\beta V(x)}$ and the shallow potential $U(x)$ is switched on at time $t=0$. Then, the system remains in the exploration phase in the shallow potential $U(x)$ for a time-interval $\mathcal{T}_1$ [drawn from the probability density function $f(\mathcal{T}_1)$]. The propagator of the system [in $U(x)$] to be at $x'$ at time $\mathcal{T}_1$ starting from $x$ is $p(x',\mathcal{T}_1|x)$. Then, the stiff/sharp trap $V(x)$ is switched on (the resetting phase) for a time-interval $\tau_1$ [drawn from $h(\tau_1)$]. In the remaining time $t-\mathcal{T}_1-\tau_1$, no resetting event occurs, and the system stays in the shallow potential $U(x)$ with the probability $g(t-\mathcal{T}_1-\tau_1)$. For this class of trajectories, the total work is the sum of the potential energy difference arising from the potential switching twice from $V(x)$ to $U(x)$ and once from $U(x)$ back to $V(x)$ [see Eq.~\eqref{eq:work}]. The contribution to the moment generating function from such trajectories is (see Fig.~\ref{fig:split}b-i)
    \begin{equation}\label{eq:psi_i}
            \psi_1^{(i)}(k, t) \equiv  \int_0^t\rmd \mathcal{T}_1I(k,\mathcal{T}_1) \int_0^{t-\mathcal{T}_1}\rmd \tau_1h(\tau_1)~\psi_0(k,t-\mathcal{T}_1-\tau_1) \ ,  
    \end{equation}
    where, for convenience, we have defined the following integral    
    \begin{align} \label{eq:intk}
     I(k, t) \equiv f(t)~\int_{-\infty}^{+\infty} \rmd x~p_{\rm eq}(x)~e^{k[U(x)-V(x)]}\nonumber\\\times \int_{-\infty}^{+\infty} \rmd x'~p(x',t|x)~e^{k[V(x')-U(x')]}\ .
    \end{align}    
    \item[ii)] The second contribution to the case of a single reset comes from trajectories where the system starts again from a Gibbs state $p_{\rm eq}(x)\propto e^{-\beta V(x)}$ with the shallow potential $U(x)$ switched on at time $t=0$. Then, the system remains in the exploration phase in the potential $U(x)$ for a time-interval $\mathcal{T}_1$ [drawn from $f(\mathcal{T}_1)$],  but once the sharp potential $V(x)$ is turned on, the particle remains in the resetting phase until the remaining time $t - \mathcal{T}_1$. The probability that the resetting phase does not end in this time is $H(t-\mathcal{T}_1)$, with $ H(t) = 1-\int_0^t~\rmd t'~h(t')$. For this class of trajectories, the total work is the sum of the potential energy difference arising from one switch from $V(x)$ to $U(x)$ and one switch from $U(x)$ back to $V(x)$ [see Eq.~\eqref{eq:work}]. The contribution to the characteristic function is (see Fig.~\ref{fig:split}b-ii)
    \begin{equation}
    \label{eq:psi_ii}
    \psi_1^{(ii)}(k, t) \equiv     \int_0^t~\rmd \mathcal{T}_1~I(k,\mathcal{T}_1)~  H(t-\mathcal{T}_1)\ . 
    \end{equation}
    \end{itemize}

In total, the contribution to the characteristic function of the $n=1$ term is the sum of $\psi_1^{(i)}(t)$~\eqref{eq:psi_i} and $\psi_1^{(ii)}(t)$~\eqref{eq:psi_ii}:
\begin{align}
\label{eq:psi1}
    \psi_1(k, t) &\equiv     \psi_1^{(i)}(k, t)+\psi_1^{(ii)}(k, t) \nonumber \\
    &= \int_0^t\rmd \mathcal{T}_1I(k,\mathcal{T}_1)\bigg[ \int_0^{t-\mathcal{T}_1}\rmd \tau_1h(\tau_1)\psi_0(k,t-\mathcal{T}_1-\tau_1)  \nonumber \\
    & +  H(t-\mathcal{T}_1)\bigg]\ .
\end{align}

    \item Similarly, for $n\geq 2$ resetting events, we can show that
    \begin{align}
    \psi_n(k, t) \equiv \int_0^t~\rmd \mathcal{T}_1~I(k,\mathcal{T}_1)\int_0^{t-\mathcal{T}_1}~\rmd \tau_1~h(\tau_1)\nonumber\\
    \times~\psi_{n-1}(k,t-\mathcal{T}_1-\tau_1) \ . \label{psin}
    \end{align}
\end{itemize} 
%
%
%
The above Eq.~\eqref{psin} is a recursive integral relation for the contribution to the characteristic function of the work with some integer number of resets. Noting that the relation is in the form of convolutions, we can 
simplify Eq.~\eqref{psin} in Laplace space as: 
\begin{align}
\tilde{\psi}_n(k,s) = \tilde{I}(k,s)~\tilde{h}(s)~\tilde{\psi}_{n-1}(k,s)\ , \label{psi-n-eqn}
\end{align}
which is valid for $n\geq 2$, and the overhead `tilde' indicates the Laplace transform of a function $M(t)$, i.e., $\widetilde{M}(s)\equiv \int_0^\infty~\rmd t~e^{-st}{M}(t)$. Solving this recursive relation~\eqref{psi-n-eqn},  we have
\begin{align}
\tilde{\psi}_n(k,s) = \bigg[\tilde{I}(k,s)~\tilde{h}(s)\bigg]^{n-1}~\tilde{\psi}_{1}(k,s)\ , \label{psi-n-eqn-lt} 
\end{align}
for $n\geq 1$. Here $\tilde \psi_1 (k,s)$ is further related to $\tilde \psi_0 (k,s)$ through taking the Laplace transform of Eq.~\eqref{eq:psi1}.

Using this result~\eqref{psi-n-eqn-lt} and the relation between $\tilde H(s)$ and $\tilde h(s)$ [i.e., $s\tilde{H}(s)  + \tilde{h}(s) = 1$], we calculate the characteristic function~\eqref{eq:genfunc_expression} in Laplace space to be
\begin{align} 
\tilde C(k,s) & = \tilde{\psi}_0(k,s) + \sum_{n=1}^{\infty}\tilde{\psi}_n(k,s) \nonumber\\
&=\dfrac{\tilde{\psi}_0(k,s) + \tilde{H}(s)~\tilde{I}(k,s)}{1- \tilde{h}(s)~\tilde{I}(k,s)}\ . \label{trap-cks}
\end{align}
This is our first main result of the paper. We have derived the Laplace transformed moment generating function of the work for a general resetting trap $V(x)$ and exploration potential $U(x)$. The result holds for any distribution $h(\tau)$ and $f(\mathcal{T})$, under the assumption that the resetting trap is always on long enough time for the system to equilibrate in the potential $V(x)$.

\subsection{Moments for Poissonian resetting}
Equation~\eqref{trap-cks} gives the characteristic function of the work in the 
Laplace space for arbitrary potentials $V(x)$ and $U(x)$ given the probability density function for the switching times $f(\mathcal{T})$ and $h(\tau)$. It is, however, cumbersome to invert the Fourier-Laplace transform to obtain the probability density function of the work, even for simple forms of $V(x),~U(x),~f(\mathcal{T}),~h(\tau)$. Nonetheless, Eq.~\eqref{trap-cks} is useful for computing the $n$-th moment of the work (in the Laplace-space) by taking the $n$-th derivative with respect to the Fourier variable $k$ and setting $k=0$.

In the following, we consider the case of exponential resetting $f(\mathcal{T}) = r~e^{- r \mathcal{T}}$  where $r$ is a constant resetting rate. Here, we find a recursive relation for the $n$-th moment of the work in the Laplace-space:
\begin{align}
    \mathcal{L}_{t\to s}\big[\langle W^{n}\rangle(t)]& = \dfrac{1}{s+r} \bigg\langle [\Delta \phi(x)]^{n} \bigg\rangle_{\rm eq} + r \tilde{H}(s)\nonumber\\
    &\times~\mathcal{L}_{t\to (s+r)}\big[\big\langle [\Delta \phi(x)-\Delta \phi(y)]^{n}\big\rangle(t)\big]\nonumber\\
    &+~r~\tilde{h}(s) \sum_{\ell =0}^n 
    \begin{pmatrix}
    n\\\ell
    \end{pmatrix} \mathcal{L}_{t\to s}\big[\langle W^{n-\ell}\rangle(t)\big]\nonumber\\
    &\times~\mathcal{L}_{t\to (s+r)}\big[\big\langle [\Delta \phi(x)-\Delta \phi(y)]^{\ell}\big\rangle(t)\big]\ ,\label{moments}
\end{align}
for the net change in the energy due to switching from the resetting to the exploration potential: $\Delta \phi(\cdot) \equiv U(\cdot) - V(\cdot)$. In Eq.~\eqref{moments}, the time-dependent averages are calculated as
\begin{align}
\langle A^n(x,y)\rangle(t)\equiv \int_{-\infty}^\infty~\rmd x~\int_{-\infty}^\infty~\rmd y~p_{\rm eq}(x)~p(y,t|x)~A^n(x,y)\ , \label{neq-avg-main}
\end{align}
while the averages $\langle \cdot \rangle_\text{eq}$ are evaluated with respect to $p_\text{eq}(x)$. Notice that when we take $r\to 0$ in Eq.~\eqref{moments}, we get
\begin{align}
    \mathcal{L}_{t\to s}\big[\langle W^{n}\rangle (t)\big] = \dfrac{1}{s} \langle [\Delta \phi(x)]^{n} \rangle_{\rm eq}\ ,
\end{align}
which is the $n$-th moment of the work done in the absence of resetting, since the only contribution when $r=0$ is from the initial preparation of the system. 
We can use  Eq.~\eqref{moments} to obtain simpler expressions for the mean and the second moment of work:
\begin{align}
    \mathcal{L}_{t\to s}[\langle W\rangle(t)] &=  \frac{s+r}{s^2 + r s(1-\tilde{h}(s))}\nonumber\\
    &\times\left(\langle \Delta \phi(x) \rangle_\text{eq} - r \mathcal{L}_{t\to s+r} [\langle \Delta \phi(y)\rangle(t) ]\right)\ , \label{mean}\\
    \mathcal{L}_{t\to s}\big[\langle W^{2}\rangle(t)]&=\dfrac{1}{s+r(1 - \tilde{h}(s))}\nonumber\\
    &\times\bigg[\big\langle [\Delta \phi (x)]^{2} \big\rangle_{\rm eq} + \dfrac{r(s+r)}{s}\nonumber\\
    &\times~\mathcal{L}_{t\to (s+r)}\big[\big\langle [\Delta \phi(x)-\Delta \phi(y)]^{2}\big\rangle(t)\big]\nonumber\\
    &+ 2 r \tilde{h}(s)~\mathcal{L}_{t\to s}\big[\langle W\rangle(t)\big]~\bigg(\big\langle \Delta \phi(x)\big\rangle_{\rm eq} \nonumber\\
    &- (s+r)\mathcal{L}_{t\to (s+r)}\big[\big\langle [\Delta \phi(y)]\big\rangle(t)\big]\bigg)\bigg]\ \label{sec-mom},
\end{align}
where $\mathcal{L}_{t\to s}\big[\langle W\rangle(t)\big]$ in Eq.~\eqref{sec-mom} is obtained from Eq.~\eqref{mean}. Higher moments can also be obtained using the recursive relation Eq.~\eqref{moments}.

\section{Harmonic traps}
\label{opt-trap}

\begin{figure*}[!t]
    \centering
    \includegraphics[width = 0.8 \textwidth]{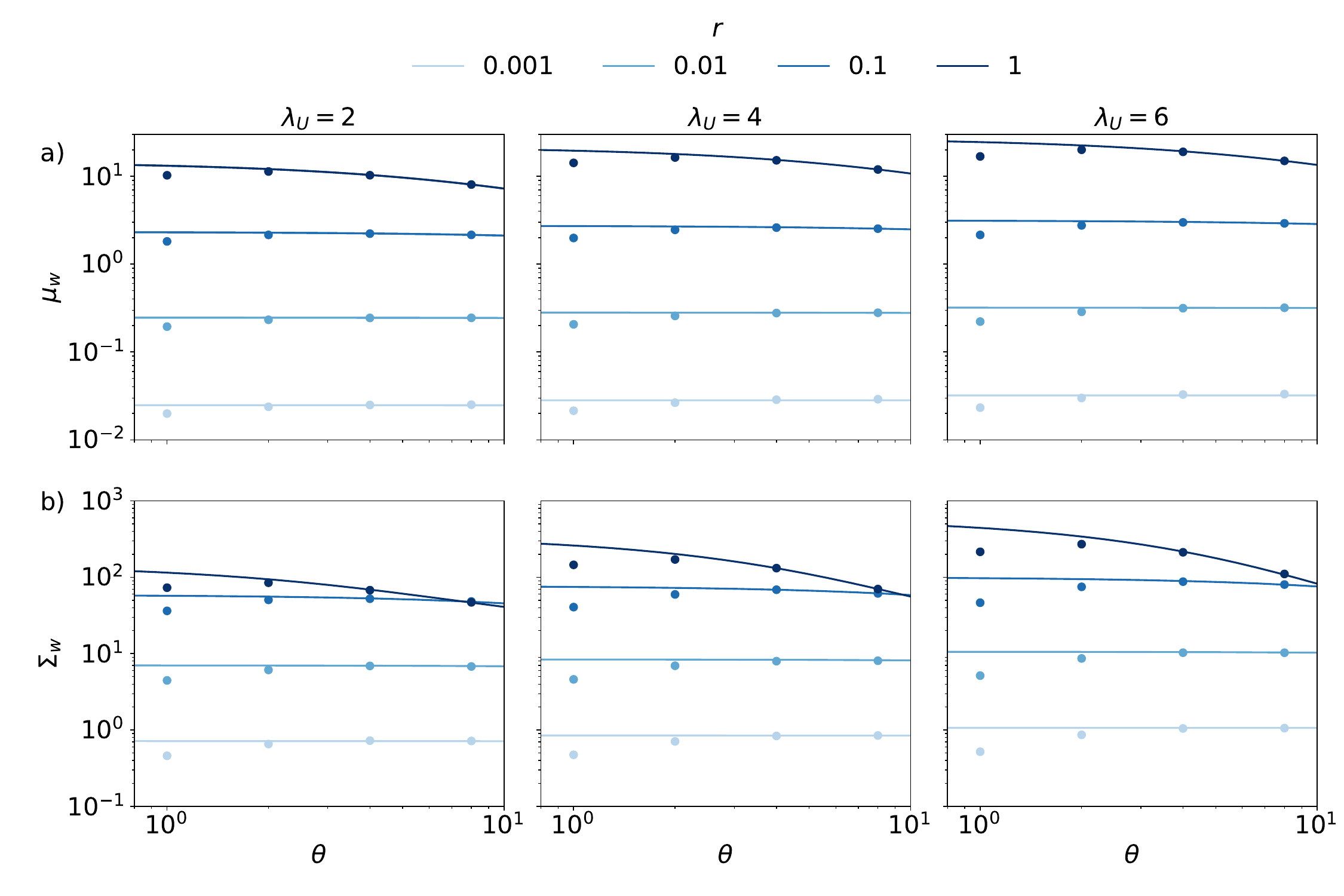}
    \caption{Long-time scaled mean ($\mu_w$) and variance ($\Sigma_w$) of work as a function of number ($\theta$) of resetting trap-relaxation time. Lines: Analytical results Eqs. \eqref{sc-mean} and \eqref{sc-var}. Symbols: Numerical simulations (without the assumption of the particle's relaxation in each return phase). The colour intensity increases with the resetting rate $r$. For numerical simulations, we take the observation time $t=10^4$. Parameters are set to $\lambda_V = 10$, $\gamma = 1$, $z = 2$ and $D = 0.5$.}
    \label{fig:sim-theory-traps}
\end{figure*}

In this section, we have a closer look at  the case of intermittent switching between two different harmonic potentials $U(x) = \frac{1}{2}\lambda_U x^2$ and $V(x) = \frac{1}{2}\lambda_V (x-z)^2$, where $z$ is the minimum of the resetting trap. For simplicity, we focus on a one-dimensional system. The general expression for the mean and variance can be  derived from 
solving Eqs.~\eqref{mean} and \eqref{sec-mom} for the case of these two harmonic potentials and for $\tilde{h}(s)=e^{-s\theta \tau_{\rm rel}}$ (which comes from our assumption that the time in the reset phase is always $\theta \tau_{\rm rel}$).   We consider durations of the exploration phase that are exponentially distributed with a constant rate $r$.

We obtain the scaled mean and variance of the work in the long-time limit by studying the Laurent expansion (around the Laplace variable $s = 0$) of Eqs.~\eqref{mean} and~\eqref{sec-mom}:
\begin{align}
 \mu_w (\lambda_U,\lambda_V)  & \equiv \lim_{t\to \infty} \frac{\langle W \rangle}{t}\ , \nonumber\\
 &= \dfrac{r}{\lambda _V(1 + r\theta \tau_{\rm rel})(r\gamma +\lambda _U)(r\gamma+2 \lambda _U)}\nonumber\\
 &\times \big[T(\lambda _U-\lambda _V)^2(r\gamma+\lambda _U)\nonumber\\
 &+z^2 \lambda_U^2 \lambda_V (r\gamma+\lambda_U+\lambda _V)\big]\ , \label{sc-mean} \\
  \Sigma_w (\lambda_U,\lambda_V)  & \equiv \lim_{t\to \infty}\frac{\langle W^2 \rangle - \langle W \rangle^2}{t}\nonumber\\
  &= \Sigma_1(\Sigma_2 + \Sigma_3 + \Sigma_4)\ ,    \label{sc-var}
\end{align}
where the resetting trap relaxation time is $\tau_{\rm rel}\equiv \gamma/\lambda_V$. The various contributions to~{\color{black}Eqs.~\eqref{sc-mean} and \eqref{sc-var}} from the general expression Eqs.~(\ref{mean}) and~(\ref{sec-mom}) are given in Appendix~\ref{sec:misc}---the small-$s$ (or, equivalently, the long-time) behaviour of these can then be extracted. The terms and factors in Eq.~(\ref{sc-var}), $\Sigma_i$, are given in Appendix~\ref{sec:var-exp}.

\begin{figure}
    \centering
    \includegraphics[width = \columnwidth]{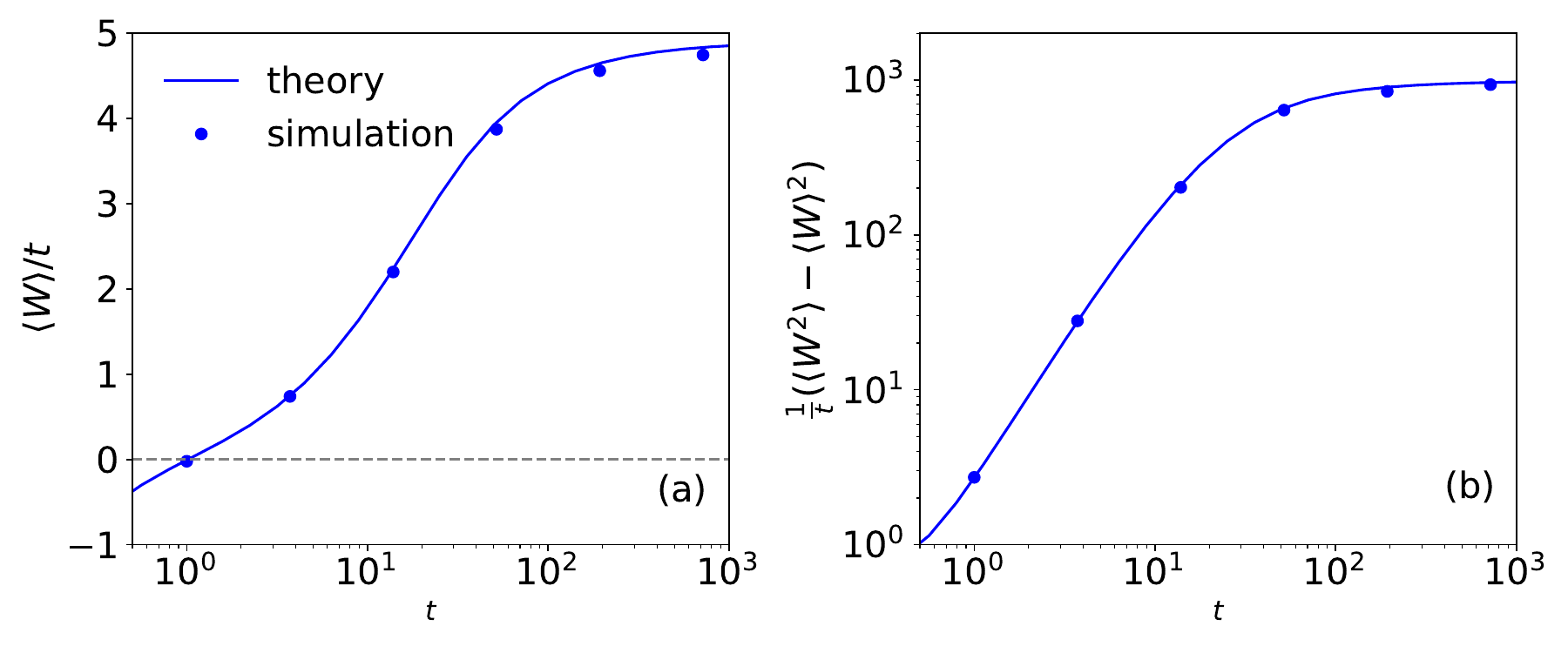}
    \caption{Free exploration case ($\lambda_U=0$). Scaled mean and variance as a function of time. Line: Theory~[obtained by numerical inversion of the Laplace transform of Eqs.~\eqref{mean} and \eqref{sec-mom}]. Symbol: Numerical simulations (without the assumption of the particle's relaxation in each return phase). We set the diffusion constant $D=0.5$, the dissipation coefficient $\gamma=1$, the reset trap's stiffness $\lambda_V = 10$, the trap's minimum at $z=2$, and number of relaxation time $\theta = 2$, resetting rate $r=0.1$. For numerical simulations, we take the time step $\Delta t = 10^{-4}$, and number of realizations $N_\mathcal{R} = 10^4$.}
    \label{fig:free-expl}
\end{figure}

The formalism developed in this work relies on the renewal structure that arises when the system relaxes to $p_{\rm eq}(x)$ after every reset. Hence, we compare our theory with numerical simulations to check when our assumptions hold. Figure~\ref{fig:sim-theory-traps} shows a comparison of analytical results~[Eqs.\eqref{sc-mean} and \eqref{sc-var}] with the long-time numerical simulations of the scaled mean and variance as a function of the number of trap-relaxation times $\theta$. Notice that in the numerical simulations (in contrast to the analytical framework in Sec.~\ref{sec:theory}) we do not enforce that the particle relaxes to the Gibbs state $p_{\rm eq}(x)\propto e^{-\beta V(x)}$ at the end of each return phase. We find that the agreement between analytical results and numerical simulations improves with increasing $\theta$, as expected, and is in fact surprisingly good even for $\theta$ between $2$ and $3$.

Figure~\ref{fig:free-expl} displays the scaled mean and variance as a function of time for arbitrary $t$ by 
numerical inversion of the Laplace transform of Eqs.~\eqref{mean} and \eqref{sec-mom} for harmonic return potential $V(x) = \frac{1}{2}\lambda_V (x-z)^2$ and free exploration phase [$U(x)=0$]. As can be seen, for the value of $\theta =2$ taken for the simulations, our theory matches the simulations for all time, indicating that the assumption of a renewal structure is valid even at moderate values of $\theta$. Notice that in the numerical simulations we do not assume the system's relaxation to the Gibbs state $p_{\rm eq}(x)\propto e^{-\beta V(x)}$ at the end of each return phase.

Note that initially the particle is in equilibrium with the potential $V(x)$. At time $t=0$, the potential is switched off, resulting in a net extraction of energy from the system [see Eq.~\eqref{eq:work}]. As a consequence, at short times, the average work is negative, as shown in Fig.~\ref{fig:free-expl}. Following additional resetting events, the average work grows, becoming positive at intermediate and longer times.

\begin{figure*}[t]
    \centering
    \includegraphics[width =  0.9 \textwidth]{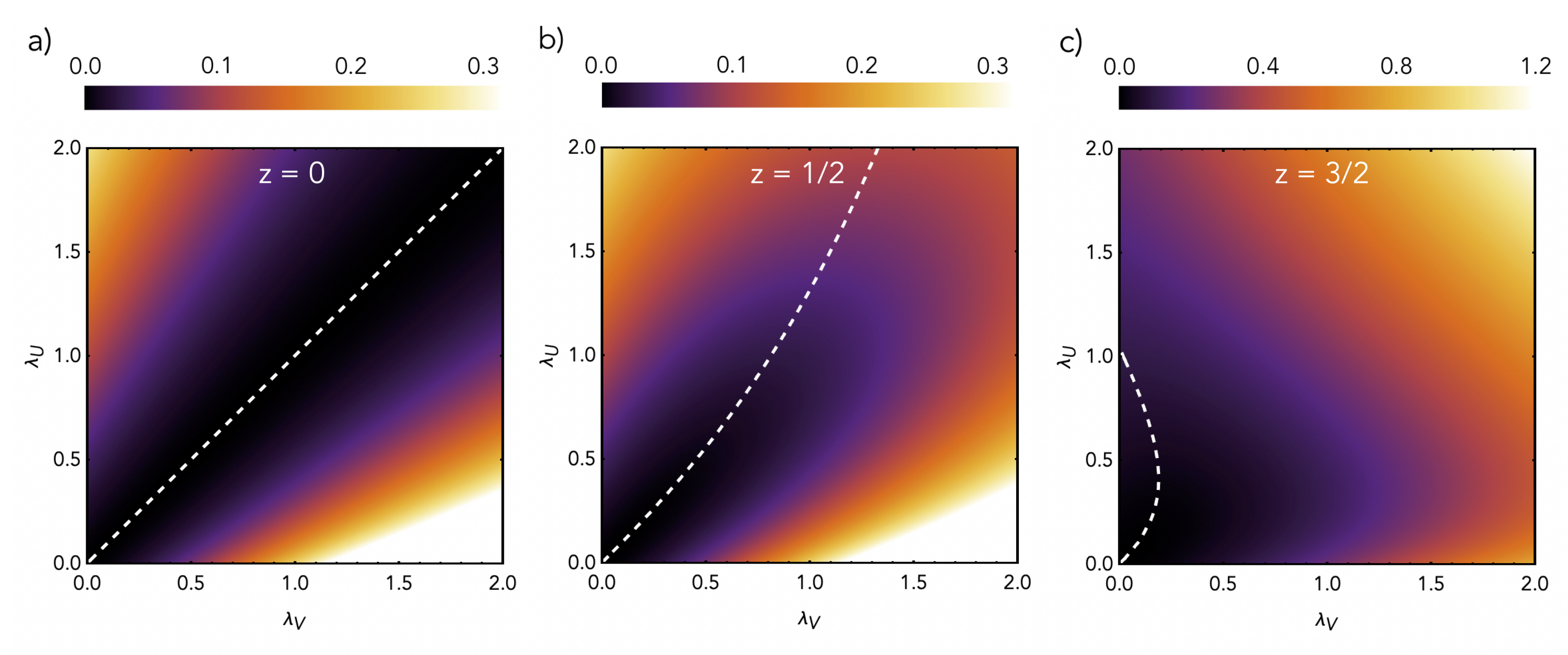}
    \caption{ Long time rate of average work Eq.~\eqref{sc-mean} as a function of $(\lambda_V,\lambda_U)$. The dashed line corresponds to the minima given by Eq. (\ref{eq:opt}), giving an optimal value of $\lambda_V^*$ for a given $\lambda_U$ and $z$. Panels a, b, and c, respectively, correspond to $z = 0,~0.5,~1.5$. Parameters are fixed to $T = r = \gamma = 1, \theta = 3$.}
    \label{fig:workmin}
\end{figure*}

For free exploration $U(x) =0$, the mean rate of work takes a particularly simple form [see Eq.~\eqref{sc-mean}]
\begin{equation}\label{eq:rateofwork}
    \mu_w (0,\lambda_V)  = \frac{D \lambda_V}{1+r \theta \tau_\text{rel}}\ ,
\end{equation}
where we used the Einstein relation $D = T/\gamma$ for the Boltzmann's constant $k_{\rm B}=1$. It is worth pointing out that this work rate does not vanish in the $r\to 0$ limit. {\color{black} This is similar to the behavior observed in Refs.~\cite{pal2023thermodynamic,Sunil_2023, tal2020experimental}. In the present case, a non-zero work as $r\to 0$ can be understood as being a result of} two competing effects: 1) At very small $r$, 
resetting events become rare, and the work goes down. However, 2) when a resetting eventually occurs, the cost associated with it is very high because particles have strayed far from the resetting location. While these two effects are always present, their cancellation at small $r$ is particular to the harmonic resetting trap for this choice of exploration duration $f(\mathcal{T}) = r \exp(-r \mathcal{T})$. Indeed, with no exploration potential, the work of a single reset in a potential $V(x)\sim |x|^q$ is expected to be proportional to $\langle w  \rangle\sim  \langle |x_\mathcal{T}|^q  \rangle$ from Eq. \eqref{eq:work}, where $\mathcal{T}$ is the time the particle has explored freely. Using the fact that the free propagator is Gaussian and averaging over this time gives $\langle w  \rangle\sim r^{-q/2}$. We see that the work for a single reset should diverge as $r\to 0$, for any potential that is confining $q>0$. The total work up to time $t$ is expected to scale as $\langle W  \rangle \sim n(r,t) \langle w \rangle$, with $n(r,t)$ the number of resets up to time $t$. For very small $r$, the exploration phase duration diverges and $n(r,t)$ resets happen in time $t \approx n/r$ as long as the resetting phase durations are large but finite. Hence, in time $t$ the work rate scales as $\langle W  \rangle /t \sim r^{1-q/2} $ for small values of $r$. Only for the harmonic case $(q=2)$ does the mean work become $r$-independent at small values of $r$.  If an exploration potential $U(x)$ is also present, all trajectories will be confined. In this case, the work vanishes for small $r$ since the work for a single reset is simply the free energy difference between the two potentials, so that $\langle W \rangle/t \sim r\Delta \mathcal{F} $.

\begin{figure*}[t]
    \centering
    \includegraphics[width =  0.94 \textwidth]{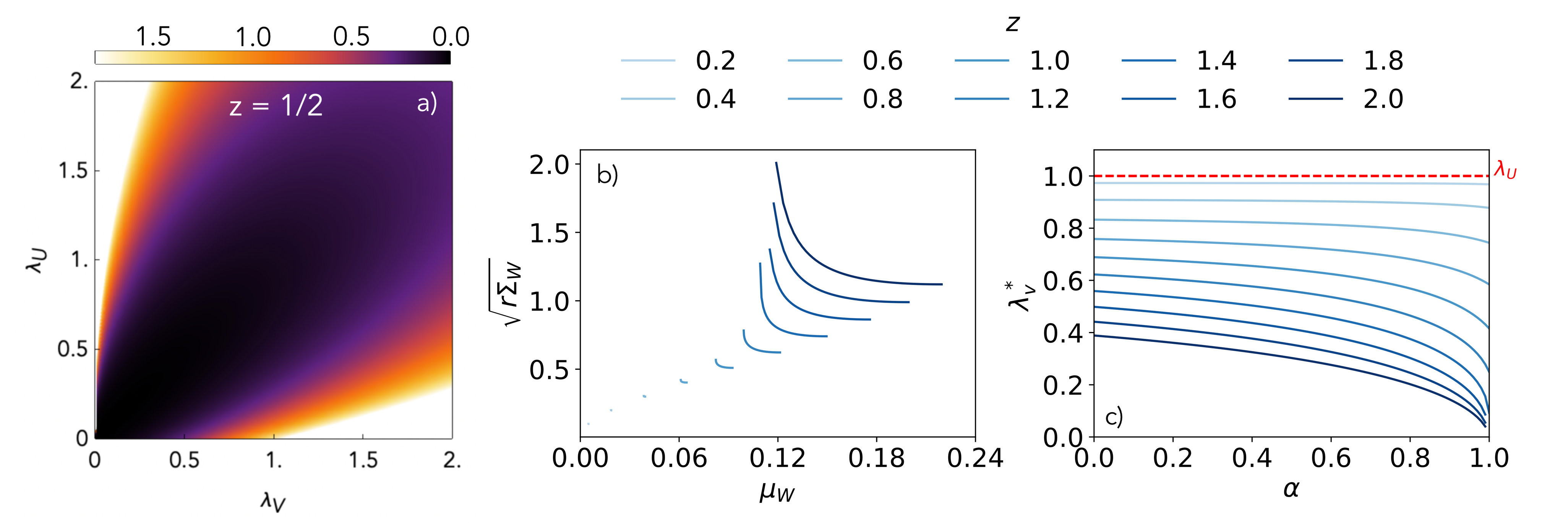}
    \caption{ a) Scaled variance $\Sigma_W$ at long times given by Eq.~\eqref{sc-var} as a function of exploration ($\lambda_U$) and resetting ($\lambda_V$) potentials'  stiffness. Resetting trap minimum is set to $z = 1/2$.
    {\color{black}b) Trade-off relation between scaled standard deviation and mean work rates, for different values of the shift parameter $z$. Each curve shows the optimal values of scaled mean and scaled standard deviation obtained by substituting $\lambda_V^*(\alpha)$~\eqref{opt-lamb} in Eqs.~\eqref{sc-mean} and~\eqref{sc-var}.} 
    c) Optimal resetting trap 
    {\color{black}strength} $\lambda_V^*$~\eqref{opt-lamb}
    as a function of the parameter $\alpha$, for different values of {\color{black}the shift parameter $z$. The red horizontal dashed line corresponds to $\lambda_U=1$ [fixed for panels b) and c)], and the color intensity increases with increasing the value of $z$. Other fixed parameters are $T = r = \gamma = 1, \theta = 3$.}}
    \label{fig:pareto}
\end{figure*}

\subsection{Optimal thermodynamic cost}
\label{otc}
For the case of Poissonian resetting at a constant rate $r$, the mean thermodynamic cost can be optimized with respect to the resetting trap $V(x)$.  We aim to identify an optimal resetting trap characterized by $\lambda_V^*$ for fixed $\lambda_U$ and $z$, so that the work performed to reset the system is minimal.

Note from Eq.~\eqref{sc-mean} that when both traps are centered at the same location ($z=0$) the mean rate of work $\mu_w (\lambda_U,\lambda_V)$ turns out to be proportional to $(\lambda_U-\lambda_V)^2$. Therefore, the work done can be globally minimized when $\lambda_U = \lambda_V$. Indeed, when $\lambda_U = \lambda_V$ and $z=0$ the two potentials coincide and hence the total work vanishes.

For $z\neq 0$, the optimal shape of the resetting trap for fixed exploration potential can be found by solving $\partial_{\lambda_V^*} \mu_w (\lambda_U,\lambda_V^*) =0$ for $\lambda_V^*$, resulting in 
\begin{align}\label{eq:opt}
\lambda_V^* =&\frac{\sqrt{\Xi  \left[\Xi (\gamma  \theta   r)^2+\lambda _U \left(\gamma 
   r+\lambda _U\right) \left\{\gamma  \theta  r \left(2 T-z^2 \lambda _U\right)+T \lambda
   _U\right\}\right]}}{\Xi }\nonumber\\
   &-\gamma  \theta  r\ ,
\end{align}
where for convenience, we have introduced $\Xi \equiv r T \gamma + \lambda_U (T + z^2 \lambda_U) $. This minimum for $z \neq 0$ is depicted in Fig.~\ref{fig:workmin} as white dashed lines. As $z \to 0$, we find $\lambda_V^* = \lambda_U$ as expected (Fig.~\ref{fig:workmin}a), reducing to the trivial case of no work performed. In the general case ($z\neq 0$), we note that the minimum is found when $\lambda_V^*<\lambda_U$, as is seen in Figs.~\ref{fig:workmin}b and~\ref{fig:workmin}c. In Fig. ~\ref{fig:workmin}c we also see that an optimal value for $\lambda_V$ only exists for a certain range of $\lambda_U$. Indeed, by solving Eq.~\eqref{eq:opt} by setting $\lambda_V^*=0$ for $\lambda_U$ in the case $z\neq 0$, we find the point where the dashed line intersects the $\lambda_U$-axis. This equation has two positive roots, with only one being non-zero. This means that there is an optimal shape of the resetting trap only when 
\begin{equation}
    0 \leq \lambda_U \leq \frac{2 r T \gamma \theta}{r\gamma \theta z^2 - T}\ . \label{eq:bound}
\end{equation}
{ If this bound~\eqref{eq:bound} is not satisfied, the optimal resetting trap shape given by Eq. (\ref{eq:opt}) does not exist, as seen by the fact that the right-hand side of Eq.~(\ref{eq:opt}) becomes imaginary.}

The {\color{black}scaled} standard deviation, $\sqrt{\Sigma_W}$, of the long-time work~\eqref{sc-var} show a similar structure to the mean in regard to its dependence on the two potentials, as in seen in Fig.~\ref{fig:pareto}a.

Often it is desirable to minimize not only the mean cost of performing resetting, but also try to keep the fluctuations to a minimum. It is, therefore, in these cases useful to introduce~\cite{pareto_1, pareto_2,pareto_3,Busiello_2021,pareto_5}
\begin{equation}\label{eq:pareto}
    \Pi_\alpha(\lambda_V) \equiv \alpha \mu_W(\lambda_V) + (1-\alpha) \sqrt{ r \Sigma_W(\lambda_V)} \ ,
\end{equation}
which considers a weighted combination of 
scaled mean and scaled standard deviation. 
The parameter $\alpha \in (0,1)$ 
is the relative weight that we want to assign to the scaled mean {\it vs.} the scaled standard deviation. Clearly, minimizing $\Pi_\alpha(\lambda_V)$ with respect to $\lambda_V$ for $\alpha = 0$ ($\alpha = 1$) minimizes the scaled standard deviation (scaled mean).

Then, for each fixed $\alpha$ we find an optimal value of trap stiffness $\lambda_V$ (for other fixed parameters) so that
\begin{align}
 \lambda_V^*(\alpha) = \underset{\lambda_V}{\rm arg~min} ~\Pi_\alpha(\lambda_V)\ ,   \label{opt-lamb}
\end{align}
simultaneously minimizes $\mu_W$ and $\sqrt{r \Sigma_W}$~\cite{pareto_1,pareto_2}. \\

Figure~\ref{fig:pareto}b displays the trade-off relation between the scaled mean and scaled standard deviation of work, i.e., the Pareto front. The weight parameter $\alpha$ parameterizes the scaled mean and scaled standard deviation by substituting $\lambda_V^*(\alpha)$~\eqref{opt-lamb} in Eqs.~\eqref{sc-mean} and~\eqref{sc-var}. Decreasing the mean rate of work is generally accompanied by an increase in its standard deviation. The trade-off region decreases with decreasing the value of $z$ as it leads to the case when the optimal trap stiffness~\eqref{opt-lamb} $\lambda_V^*\to \lambda_U$ for all values of $\alpha$~(Fig.~\ref{fig:pareto}c). This is expected because as $z\to 0$ and $\lambda_V\to \lambda_U$, both scaled mean~\eqref{sc-mean} and scaled standard variance~\eqref{sc-var} approach zero. This is due to the fact that no work will be performed in this case [see Eq.~\eqref{eq:work}].

\section{Mean work under arbitrary durations of the two phases}
\label{sec:finite}
In the above sections, we assumed that the resetting trap is active for a sufficiently long time so that the particle relaxes to a Boltzmann state at the end of each resetting phase, i.e., the system is assumed to reset according to $p_{\rm eq}(x)\propto e^{-\beta V(x)}$. However, certain experimental situations may not allow this strict condition. In order to circumvent such issues, in this section, we  relax this assumption and derive a general expression for the mean work for arbitrarily distributed durations of both resetting and exploration phases. 

For a fixed observation time and arbitrary resetting $V(x)$ and exploration $U(x)$ potentials, this calculation becomes cumbersome. However, it turns out that one can still find the mean work for a fixed number $n$ of {\it return trails} with the exploration phase consisting of free diffusion $U(x)=0$, and resetting facilitated by a harmonic potential $V(x) = \lambda_V x^2/2$. (This situation can be easily realized in experiments~\cite{besga2020optimal, goerlich2023experimental}.) Note that each return trail is composed of an exploration phase, followed by a resetting phase. Then, following Eq.~\eqref{eq:work} the net average work done after exactly $n$ return trails is
\begin{equation}  
    \label{eq:meanwork_general}
    \langle W_n \rangle =  \dfrac{\lambda_V}{2}\sum_{j=1}^n [\langle(x_{2j-1}^+)^2\rangle-\langle(x_{2j}^-)^2\rangle]\ ,
\end{equation}
where the angular brackets indicate the average performed over the probability density function of the particle to be at $x_{2j-1}^+$ and $x_{2j}^-$. We stress that these positions are the extreme points on the upper horizontal line in Fig.~\ref{fig:scheme}b-c for each return duration $t_{2j}-t_{2j-1}$. Further, for convenience, we henceforth drop the superscripts $`\pm'$ from $x^\pm_{\dots}$. Here, we do not make any assumptions regarding the initialization of the system, so we do not include a cost associated with its preparation at $t=0$.

To proceed, we find the probability density function for the particle to be at $x_{2j-1}$ and $x_{2j}$, which we respectively denote as { $\rho_{\rm E}^{({2j-1})}(x_{2j-1}|x_0)$}  and { $\rho_{\rm R}^{({2j})}(x_{2j}|x_0)$}. Here, { the $E$ and $R$ subscripts label the density for the particle's position at the end of the exploration and resetting phase respectively. }
We construct these distributions iteratively by noting that the probability of the particle to be at $x_{2j-1}$ given that it started from $x_0$, is obtained by considering the contribution of trajectories reaching $x_{2j-2}$ starting from $x_0$ [{ governed by $\rho_{\rm R}^{({2j-2})}(x_{2j-2}|x_0)$}], after which the system performs free diffusion until a duration of one reset interval drawn from $f(\mathcal{T})$. A similar logic can be applied to the probability of finding the particle at $x_{2j}$. Then, we have the recursion relations{
\begin{align}
    \rho_{\rm E}^{({2j-1})}(x_{2j-1}|x_0) &=  \int_{-\infty}^{+\infty}~\rmd x'_{2j-2}~\int_0^\infty~\rmd \mathcal{T}~f(\mathcal{T})\nonumber\\
    &\times~p_0(x_{2j-1},\mathcal{T}|x'_{2j-2})~\rho_{\rm R}^{({2j-2})}(x'_{2j-2}|x_0)\ , \label{eq:h1}\\
    \rho_{\rm R}^{({2j})}(x_{2j}|x_0) &=  \int_{-\infty}^{+\infty}~\rmd x'_{2j-1}~\int_0^\infty~\rmd \tau~h(\tau)\nonumber\\
    &\times~p_V(x_{2j},\tau|x'_{2j-1})~\rho_{\rm E}^{({2j-1})}(x'_{2j-1}|x_0)\ , \label{eq:h2}
\end{align}}
for {$1\leq j\leq n$ and $\rho^{(0)}_{\rm R}(x'_{0}|x_0) = \delta(x_0'-x_0)$}.  For simplicity, we focus on a one-dimensional system. In Eqs.~\eqref{eq:h1} and~\eqref{eq:h2}, the probability density functions corresponding to the free diffusion and diffusion in the harmonic potential, respectively, are
\begin{align}
    p_0(x,t|x_0) &= \dfrac{e^{-\frac{-(x-x_0)^2}{4 D t}} }{\sqrt{4 \pi D t}} \ ,\\
    p_V(x,t|x_0) &= \dfrac{e^{-\frac{-(x-\mu(t))^2}{2 \sigma^2(t)}}}{\sqrt{2 \pi \sigma^2(t)}} \ ,
\end{align}
for mean $\mu(t) \equiv x_0~e^{-t/\tau_{\rm rel}}$ and variance $\sigma^2(t) \equiv D \tau_{\rm rel}(1-e^{-2 t/\tau_{\rm rel}})$, where we defined the trap relaxation time $\tau_{\rm rel} \equiv \gamma/\lambda_V$.

Using Eq.~\eqref{eq:h1}, we calculate the average $\langle x_{2j-1}^2 \rangle$ as follows:
\begin{align}
    \langle x_{2j-1}^2 \rangle &=   \int_{-\infty}^{+\infty}~\rmd x'_{2j-2}~\int_0^\infty~\rmd \mathcal{T}~f(\mathcal{T})~{\rho_{\rm R}^{({2j-2})}(x'_{2j-2}|x_0)} \nonumber\\
    &\times\left[ 2 D \mathcal{T} + (x'_{2j-2})^2 \right]\  \nonumber\\
    &=  2 D \langle \mathcal{T} \rangle +  \langle x_{2j-2}^2 \rangle\ ,\label{eq:univ1}
\end{align}
where $\langle \mathcal{T} \rangle \equiv \int_0^\infty~\rmd t~t f(t)$ is the mean time-interval of the exploration phase.

\begin{figure}
    \centering
    \includegraphics[width =  7.5 cm]{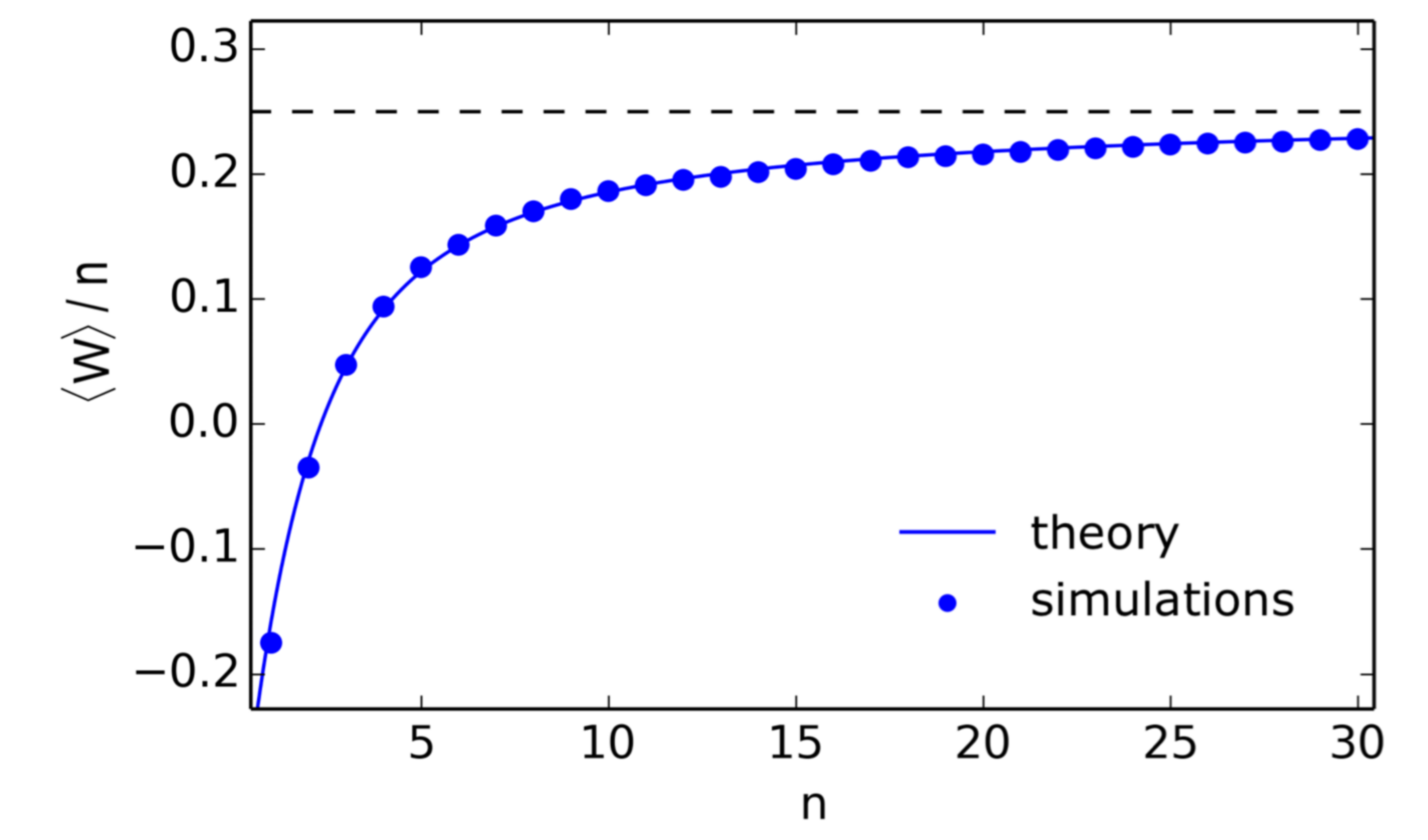}
    \caption{Mean work rate $\langle W\rangle /n$ of resetting as a function of number $n$ of resetting trails, for exponentially distributed exploration phases with rate $r$ and  resetting phases of fixed duration $\theta \tau_{\rm{rel}}$. Solid line: Eq.~\eqref{eq:mainres}. Symbols: Numerical simulation. Number of realizations $N_R = 10^4$. The black dashed line shows the steady-state rate $\lambda_V D/r$~\eqref{large-n}. Parameters used are $r = D = \gamma = 1, \lambda_V = 0.25, \theta = 0.5, x_0 = 0$. }
    \label{fig:workneg}
\end{figure}

Next, we derive a hierarchy for $ \langle x_{2j}^2 \rangle$, which upon solving requires $ \langle x_{2j-1}^2 \rangle$:
\begin{align}
    \langle x_{2j}^2 \rangle  = D \tau_{\rm rel} \left\langle 1 - e^{-2 \tau/\tau_{\rm rel}} \right\rangle + \left\langle  e^{-2\tau/\tau_{\rm rel}} \right\rangle
    \langle x_{2j-1}^2 \rangle\ . \label{x_even}
\end{align}
Here the average is performed over the distribution of resetting trap switch-on time interval, i.e., $h(\tau)$. One should note that in obtaining Eq. (\ref{x_even}) we made use of the second moment of the propagator in the harmonic potential. We emphasize that generalizations to other resetting traps enter at this stage.  The above equation~\eqref{x_even} simplifies to  
\begin{align}
    \langle x_{2j}^2 \rangle = D \tau_{\rm rel} \big[1 - \tilde{h}(2/\tau_{\rm rel})\big]+ \tilde{h}(2/\tau_{\rm rel})~\langle x_{2j-1}^2\rangle\ ,\label{eq:univ2}
\end{align}
where the tildes denote Laplace transformations. Equations ~\eqref{eq:univ1} and~\eqref{eq:univ2} can be solved recursively. Combined with  Eq.~\eqref{eq:meanwork_general}, we get the average work performed to reset a system after $n$ return trails:
\begin{align}\label{eq:mainres}
     \left \langle W_n \right\rangle =& \lambda_V\bigg[D\langle \mathcal{T} \rangle \bigg\{n - \tilde h(2\tau_{\rm rel}^{-1}) \dfrac{1-\tilde{h}^n(2\tau_{\rm rel}^{-1})}{1-\tilde h(2\tau_{\rm rel}^{-1})}\bigg\} \nonumber\\
     &+ \dfrac{1 - \tilde{h}^n(2\tau_{\rm rel}^{-1})}{2}\big(\langle x_0^2 \rangle - D \tau_{\rm rel}\big) \bigg] \ . 
\end{align}
This analytical result is compared with the numerical simulations in Fig.~\ref{fig:workneg}, and it shows an excellent agreement.

A couple of interesting remarks are in order. First, consider the case when the resetting trap is switched on for a fixed duration, i.e., $h(\tau) = \delta(\tau-\theta \tau_{\rm rel})$. Then, the limit $\theta \to 0$ implies that the resetting potential, $V(x)$, is never turned on. This gives $\tilde h(2\tau_{\rm rel}^{-1}) \to 1$ in the above expressions. This limit must be taken with some care. Using L'H\^{o}pital's rule, we can show that the average work $\langle W_n \rangle= 0$ as expected, since the potential is never turned on.

Secondly, the mean work may be negative for small values of $n$. For example, after one return trial, we find the work 
\begin{equation}
     \left \langle W_1 \right\rangle = \dfrac{1}{2}\lambda_V\left[1 - \tilde{h}(2\tau_{\rm rel}^{-1})\right]\left [\langle x_0^2 \rangle + D(2\langle \mathcal{T}\rangle - \tau_{\rm rel})\right]\ .
     \label{eq:w1}
\end{equation}
Since  $\hat h(2 \tau_\text{rel}^{-1}) < 1$, the work after one return trail is negative if the expression inside the right-most square brackets is negative; namely when
\begin{equation}\label{eq:ineq}
    \langle \mathcal{T}\rangle < \dfrac{\tau_{\rm rel} - \langle x_0^2 \rangle/D}{2}\ .
\end{equation}
An example of this is seen in Fig.~\ref{fig:workneg}. If the particle is initialized in a Boltzmann state of the resetting trap $V(x)$, $\langle x_0^2 \rangle = D \tau_\text{rel} $ and one would require negative $ \langle \mathcal{T} \rangle$ to obtain negative work, which of course can never happen. Hence, negative work after one reset cycle requires some fine-tuning of the initial position.  This is simply because after a careful initialization near the potential minima, any thermal fluctuation is likely to bring the particle to a higher energy state at the end of one resetting trail.

The above results are general and hold for any distribution of durations for the resetting and exploration phases. In contrast, Sec.~\ref{sec:gen-func} assumes that the duration of the resetting phase is sufficiently large so that the system relaxes to a Boltzmann state corresponding to the resetting trap $V(x)$. Interestingly, for large $n$ (or equivalently at longer times) the average work done Eq.~\eqref{eq:mainres} behaves as 
\begin{align}
     \left \langle W_n \right\rangle \to & \: n D \lambda_V  \langle \mathcal{T} \rangle \ . \label{large-n}
\end{align}

In this section, we have considered an ensemble of paths for which the number of resets is fixed while the observation time can fluctuate. The mean observation time during $n$ resets is therefore $\langle { t_n} \rangle = n [\langle \mathcal{T} \rangle+\langle \tau \rangle]$. Hence, we can express the { asymptotic} mean work Eq.~\eqref{large-n} rate for the mean time $\langle {t_n} \rangle$ as
\begin{align}\label{eq:finitetime2}
     \frac{\left \langle  W_n \right\rangle}{ \langle { t_n} \rangle} \approx &  \frac{D\lambda_V }{1 + \langle \tau\rangle / \langle \mathcal{T} \rangle }\ . 
\end{align}
Interestingly, this closely resembles the calculations where it was assumed that $\tau$ was much larger than the relaxation time of the resetting trap, resulting in the system relaxing to a Boltzmann state after each reset. See for example Eq.~\eqref{eq:rateofwork} with $\langle \mathcal{T}\rangle = 1/r$ and $\langle \tau \rangle = \theta \tau_{\rm rel}$. 

\subsection{Average work for two harmonic potentials}\label{sec:SI_twopots}

Here, we extend the previous calculation to the case of intermittent switching between \emph{two} harmonic potentials. Rather than considering the full solution at any $n$, we restrict our attention in this case to the long-time behavior. We consider potentials  (exploration phase) $U(x) = \frac{1}{2}\lambda_U x^2$ and (return phase) $V(x) = \frac{1}{2}\lambda_V x^2$.  In the late-time (large $n$) limit, we expect that the average work in Eq. \eqref{eq:work} grows as
\begin{align}
    \langle W_n \rangle \approx  n \frac{\lambda_V - \lambda_U}{2}  [\langle x_E^2\rangle-\langle x_R^2\rangle]\ , \label{mean-new}
\end{align} 
where $x_{E,R}$ denotes the positions in the steady state right after the exploration phase (E) and return phase (R). Here we have simply used the fact that for large $j$ in Eq.~\eqref{eq:work}, as the steady state is approached, both second moments approach their steady-state values. While this is an approximation for finite $n$, we expect it to become exact in the limit $n \to \infty$. 

As before, the second moments can be obtained using Eqs. \eqref{eq:h1} and \eqref{eq:h2}, which at large $j$ give the following set of coupled steady-state equations
\begin{align}
    \langle x_E^2\rangle &=D\lambda_U^{-1}+[\langle x_R^2\rangle-D\lambda_V^{-1}]\tilde{h}(2\lambda_V)\ , \label{eq125} \\
    \langle x_R^2\rangle  &=D\lambda_V^{-1}+[\langle x_E^2\rangle-D\lambda_V^{-1}]\tilde{f}(2\lambda_E)\ . \label{eq126}
\end{align}
Solving above equations~\eqref{eq125} and \eqref{eq126} gives the steady state moments at the end of the exploration and return phases as
\begin{align}
    \langle x_E^2\rangle &= \frac{ D\lambda_U^{-1}[1-\tilde{f}(2\lambda_U)]+D\lambda_V^{-1}[1-\tilde{h}(2\lambda_V)]\tilde{f}(2\lambda_U)}{1-\tilde{f}(2\lambda_U)~\tilde{h}(2\lambda_V)}\ , \\
    \langle x_R^2\rangle &= \frac{ D\lambda_V^{-1}[1-\tilde{h}(2\lambda_V)]+D\lambda_U^{-1}[1-\tilde{f}(2\lambda_U)]~\tilde{h}(2\lambda_V)}{1-\tilde{f}(2\lambda_U)~\tilde{h}(2\lambda_V)}\ .
\end{align}

\begin{figure}
    \centering
    \includegraphics[width =  8.5 cm]{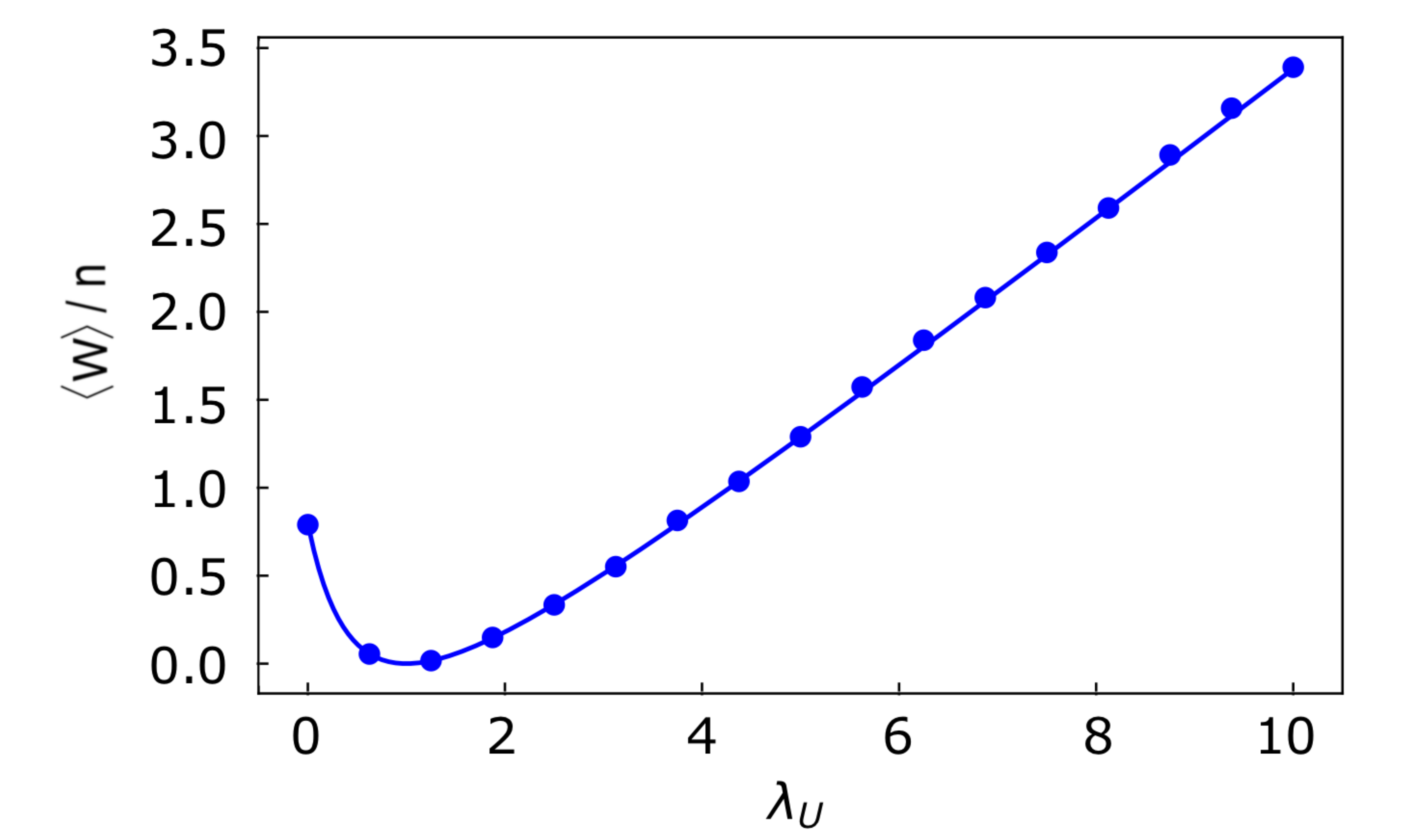}
    \caption{Mean work rate $\langle W\rangle /n$ for large $n$ of resetting as a function of the exploration potential's strength $\lambda_U$, for $h(\tau)=\delta(\tau-1)$ and $f(\mathcal{T})=e^{-\mathcal{T}^2/2}\sqrt{2/\pi}$ with $\mathcal{T}>0$. Solid line: Exact result~\eqref{eq:work_gen_example}. Symbols: Numerical simulation.  Each data point is obtained from a single trajectory of duration $t=10^7$. Parameters used are $D = \gamma = 1, \lambda_V = 1, x_0 = 0$.}
    \label{fig:workneg_si}
\end{figure}

Then, substituting $\langle x_{E, R}^2\rangle$ in Eq.~\eqref{mean-new} gives the mean rate of work as follows
\begin{equation}
 \lim_{n\to \infty}   \frac{\langle {W_n}\rangle}{n}= \frac{D}{2} (\lambda_V-\lambda_U)^2
 \frac{[1-\tilde{f}(2\lambda_U)][1-\tilde{h}(2\lambda_V)]}{\lambda_U\lambda_V[1-\tilde{f}(2\lambda_U)~\tilde{h}(2\lambda_V)]}\ .\label{eq:avg_work__gen}
\end{equation}
In the $\lambda_U \to 0$ limit, where we use $\tilde f(2\lambda_U) \approx 1 - 2\lambda_U \langle \mathcal{T} \rangle$, we regain the previous expression in Eq. (\ref{large-n}). This extends previous theoretical work on intermittent and fluctuating potentials, where only exponentially distributed durations of the two phases were considered \cite{alston2022non}.

Now let us consider, as a second example, the case where the duration of the resetting phase is deterministic with $h(\tau)=\delta(\tau-\tau_{\rm on})$, and the exploration phases are distributed according to a half-Gaussian distribution $f(\mathcal{T})=e^{-\mathcal{T}^2/(2\sigma^2)}\sqrt{2/(\pi\sigma^2)}$ with $\mathcal{T}>0$. Then, using Eq.~\eqref{eq:avg_work__gen}, we find
\begin{align}
    \lim_{n\to \infty}   \frac{\langle {W_n}\rangle}{n}&= \dfrac{D}{2} (\lambda_V-\lambda_U)^2\nonumber\\
    &\times\dfrac{(e^{2\lambda_V\tau_{\rm on}}-1)[e^{2\lambda_U^2\sigma^2}\operatorname{Erfc}(\sqrt{2}\lambda_U\sigma)-1]}{\lambda_U\lambda_V [e^{2\lambda_U^2\sigma^2}\operatorname{Erfc}(\sqrt{2}\lambda_U\sigma)-e^{2\lambda_V\tau_{\rm on}}]}\ ,\label{eq:work_gen_example}
\end{align}
where $\text{Erfc}(u) \equiv \frac{2}{\sqrt{\pi}}\int_u^\infty~d t~e^{-t^2} $ is the complementary error function.
This exact expression is shown in Fig.~\ref{fig:workneg_si} and is in perfect agreement with numerical simulations.

\section{Comparisons with other resetting protocols}\label{sec:comp}

In the conventional paradigm where each reset instantaneously brings the particle position back to some initial fixed location $x_0$, it can be hard to draw conclusions regarding the behavior of thermodynamic variables. As shown recently in \cite{mori2023entropy}, a slightly redefined resetting mechanism which considers instead
instantaneous resetting to a \textit{distribution} of resetting positions makes it possible to compute the full distribution of entropy production. In the following, we refer to this resetting protocol as \textit{instantaneous distributed resetting} (IDR).

If the physical mechanism of implementing the reset by switching on a potential is taken into consideration, then the cost of its implementation can be taken into account in different ways, all of which however being of necessity non-instantaneous resetting.
In \cite{Deepak2022_work} the cost function we consider here is introduced for a very similar model. However, the difference lies in the time spent in the return phase, which in \cite{Deepak2022_work} is determined by the first passage time to the potential minimum. We refer to this protocol as \textit{first-passage resetting} (FPR).

In this section, we compare our results with previous work on the mean of the 
cost of resetting for the special case of free diffusion in the exploration phase [$U(x)=0$] and  
two generic resetting potentials: 1) the V-shaped potential $V(x) =  \gamma_V \lvert x \rvert$ and 2) the harmonic potential $V(x) = \frac{1}{2}\lambda_V x^2$. The two earlier works we compare with are the FPR~\cite{Deepak2022_work} and  IDR protocols~\cite{mori2023entropy}.

\subsection{Connection with instantaneous distributed resets}
\label{sec:connect}
In a recent work, the entropy production under instantaneous distributed resetting  was considered, where the resetting position is picked from $p_R(x)$ \cite{mori2023entropy}. As discussed therein, a physical interpretation of $p_R(x)$ is as a distribution of resetting positions that results from the application of a resetting potential $V(x)$. 
The choice $p_R(x) \sim e^{-\beta V(x)} $ is then
equivalent to our assumption here that the particle is allowed to relax to the equilibrium distribution at the end of each resetting phase. One should note that while the Ref.~\cite{mori2023entropy} considers the late-time mean medium entropy production, and we consider the mean work, the two are simply related by a factor of temperature. We denote $\mu_W^{(\text{IDR})}$ the resulting mean rate of work that results from the IDR protocol.  

To compare the current resetting protocol with the IDR protocol, one has to appreciate the different ways in which instantaneous resetting can be interpreted. The literal interpretation is that resetting actually occurs in an instant, for example achieved by instantly activating an infinitely strong potential. This would make the resetting both perfectly accurate and instantaneous. However, since the resetting trap that results in a finite-width distribution $p_R(x)$ by necessity has finite stiffness, the resetting should also take a non-zero time. This is because the resetting trap must be active for at least one relaxation time characteristic to the potential, which scales as $\lambda_V^{-1}$.

Hence, a more natural interpretation of the IDR scheme is one in which the resetting phases are removed from the data \textit{ad hoc}. This is similar to what is performed in recent experiments, where instantaneous resets are simulated by utilizing a finite time resetting scheme followed by the removal of the resetting periods \cite{goerlich2023experimental}. Due to the removal of these phases, the resulting time-series is of course shortened, and the correcting factor mapping the real observation time to the simulated observation time with instant resets is the fraction of time spent in the exploration phase
\begin{equation}\label{eq:timerescaling}
    t \longrightarrow \frac{1/r}{1/r + \langle \tau\rangle} \:\: t\ .
\end{equation}
where $1/r$ is the mean exploration time.

Hence, rescaling time accordingly, we find that the rates of work in the finite-time and instantaneous schemes are related by
\begin{equation}
    \mu_W =  \frac{1/r}{1/r + \langle \tau\rangle} \mu_W^{(\text{IDR})} =  \frac{1}{ 1 +  r \langle \tau\rangle} \mu_W^{(\text{IDR})} \label{scaling-rel}
\end{equation}
This relation can be confirmed by considering concrete examples.

\bigskip

\textbf{Harmonic traps: } In Sec.~\ref{opt-trap}, we use Eqs.~\eqref{mean} and \eqref{sec-mom} to derive a general expression for the mean work done for resetting a particle in a harmonic trap $U$ by turning on a resetting potential $V$. In the case when $U=0$, the scaled mean work~\eqref{sc-mean} is
\begin{align}
 \mu_W = \lim_{t\to \infty}\dfrac{\langle W\rangle}{t} = \dfrac{D \lambda_V}{1+r \theta \tau_{\rm rel}}\ , \label{wdot-2}  
\end{align}
where $\lambda_V$ is the stiffness of the harmonic potential.  In the case of \cite{mori2023entropy}, the total entropy production rate is 
\begin{align}
    \lim_{t\to \infty}\dfrac{\langle S_{\rm tot}\rangle}{t} = \dfrac{D}{\sigma^2}\ ,
\end{align}
for the case when $p_R(x)$ is Gaussian
\begin{align}
    p_{\rm R}(x) \equiv \frac{1}{\sqrt{2\pi\sigma^2}}e^{-x^2/(2 \sigma^2)}\ .
\end{align}
To match in notation, we must use the relation $\sigma^2 = 1/(\beta \lambda_V)$ so that $p_R(x) \sim e^{-\beta V(s)}$. We also note that to obtain the rate of work from the entropy production, we must multiply with a factor of temperature
\begin{equation}
    \mu_W^{(\text{IDR})} = T \lim_{t\to \infty}\dfrac{\langle S_{\rm tot}\rangle}{t}\ . 
\end{equation}
Combining the above, we find
\begin{equation}\label{eq:rel_harminoc}
    \mu_W =  \frac{1}{ 1 +  r \theta \tau_{\rm rel}} \mu_W^{(\text{IDR})}\ ,
\end{equation}
as expected. We emphasize again that in order to obtain equality $ \mu_W =  \mu_W^{(\text{IDR})}$ one must either consider very strong resetting traps with $r \theta \tau_{\rm rel} \ll 1$, or consider the time rescaling as in Eq.~\eqref{eq:timerescaling}.

\bigskip

\textbf{V-shaped traps:} To confirm that Eq.~\eqref{scaling-rel} is not a coincidence specific to the case of harmonic potentials, we consider also a V-shaped resetting trap $V(x) = \gamma_V |x|$. For the case of exponential resetting, we can use our results from 
Eq.~\eqref{mean} to obtain  the scaled mean work at long-time as 
(see Appendix~\ref{sec:Vcalc} for the detailed derivation):
\begin{align}\label{eq:work_V}
     \mu_W&=\dfrac{\beta D \sqrt{r} \gamma_V^2 }{\left(\sqrt{r}+\gamma_V\beta \sqrt{D} \right)(1+ r\theta\tau_{\rm rel}) }\ .
\end{align}
Here the relaxation time is a characteristic timescale proportional to $\gamma_V^{-1}$. The exact choice of $\tau_\text{rel}$ is not crucial, as prefactors can be absorbed into $\theta$. In the IDR protocol considered in \cite{mori2023entropy}, this potential corresponds to a Laplace distribution $p_R(x) \sim e^{- |x|/a}$. The rate of work (again obtained by scaling the entropy production with temperature) reads
\begin{equation}
    \mu_W^{(\text{IDR})} = \frac{T D r}{a (ar + \sqrt{D r})}
\end{equation}
In order to match notation, we set $\gamma_V = 1/(\beta a)$. We also use $k_B = \gamma = 1$, in which the Einstein relation simply reads $\beta = 1/D$. Combining the above, we again find
\begin{equation}
    \mu_W =  \frac{1}{ 1 +  r \theta \tau_{\rm rel} } \mu_W^{(\text{IDR})}.
\end{equation}

This shows that the instantaneous distributed resetting framework in \cite{mori2023entropy} can be useful as a proxy for calculating the mean rate of work or other thermodynamic quantities, with a time rescaling needed to make connections to finite-time schemes, as predicted by Eqs.~\eqref{eq:timerescaling} and \eqref{scaling-rel}.

\subsection{Comparison to first-passage resets}
The instantaneous distributed resetting scheme has a close connection to the current implementation, as they both result in particles with resetting positions distributed according to Boltzmann weights. In the first-passage resetting scheme as considered in~\cite{Deepak2022_work}, the connection is less clear.

\bigskip

\begin{figure}
    \centering
    \includegraphics[width = 0.4\textwidth]{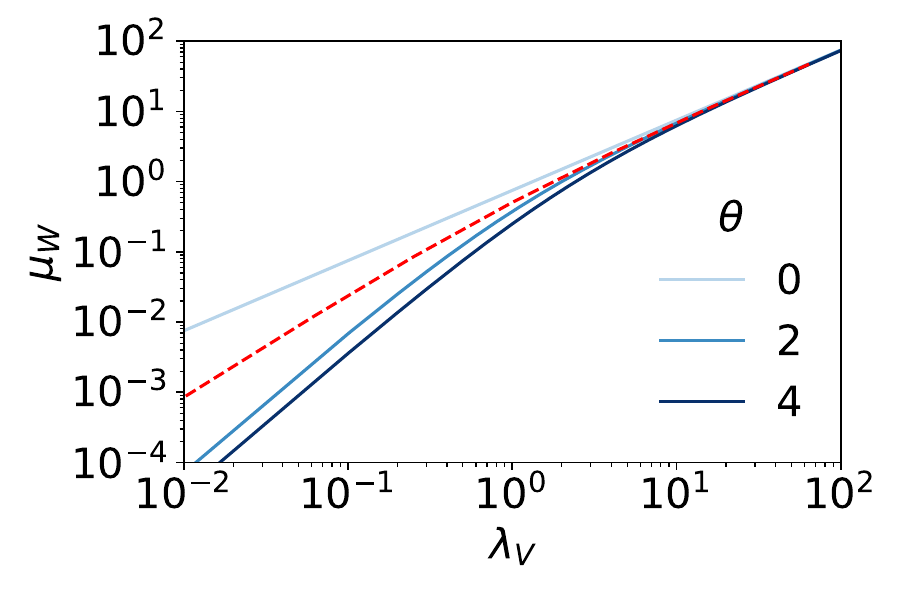}
    \caption{Harmonic return potential and free diffusion. Comparison of scaled mean work $\mu_W$~\eqref{wdot-2} (blue lines) with that of obtained in Ref.~\cite{Deepak2022_work}~[Fig. 4a(right panel)] (red dashed line), as a function of trap stiffness $\lambda_V$. The color intensity increases with increasing $\theta$. Parameters: Resetting rate $r = 0.5$, diffusion constant $D = 0.75$, and dissipation constant $\gamma = 1$.}
    \label{fig:njp-curr-manu}
\end{figure}

\textbf{Harmonic traps:} Figure~\ref{fig:njp-curr-manu} compares the scaled work \eqref{wdot-2} with the expression derived for first-passage resetting in \cite{Deepak2022_work}. We see that neither at very short nor at very long durations of the resetting phases does the rate of work match the first-passage scheme. Matching is only achieved in the large $\lambda_V$ limit when the relaxation time in the trap is much faster than the interval between resets. \\

\textbf{V-shaped traps:}  In \cite{Deepak2022_work}, the authors use parameters so that $\beta = 1/D$ as before, and additionally introduce the parameter $\alpha_V \equiv \gamma^2_V/(4 D)$. They find [see Eq.~(24) in \cite{Deepak2022_work}] the rate of work to be
\begin{equation}
   \mu_W^{(\text{FPR})} = {\dfrac{4D \sqrt{r} \alpha_V }{\left(\sqrt{r}+2\sqrt{\alpha_V} \right) }}\ .
\end{equation}
We can rewrite the work obtained using the current method, Eq.~\eqref{eq:work_V}, as 
\begin{align}
\label{mean_V}
     \mu_W&=\dfrac{1}{1+ r\theta\tau_{\rm rel}} {\dfrac{4D \sqrt{r} \alpha_V }{\left(\sqrt{r}+2\sqrt{\alpha_V} \right) }} =  \dfrac{1}{1+ r\theta\tau_{\rm rel}} \mu_W^{(\text{FPR})} \ .
\end{align}

We should stress that in contrast to the case of instantaneous distributed resetting (IDR), the first-passage resetting (FPR) scheme is not simply connected to our current scheme by a rescaling in time. Hence, Eq. (\ref{mean_V}) matches the rate of work obtained in \cite{Deepak2022_work} (FPR case) only for $r \theta \tau_{\rm rel} \ll 1$, i.e., when the time interval between resets is always much larger than the time for relaxation in the potential{, i.e., for a very stiff resetting trap.
This can be understood as follows. Since both the relaxation time and the first passage time to the trap minimum decreases with increasing trap stiffness, we expect that for stronger resetting traps, the resetting-phase duration in the finite-time protocol will be of the same order as that of the first-passage time in the first-passage protocol. Furthermore, increasing the stiffness restricts the particle in the finite-time resetting protocol to be around the resetting location (and the magnitude of positional fluctuations around this resetting location decreases as stiffness increases), and therefore, the contribution to the work from fluctuations near the resetting location will be vanishingly small (notice that this contribution is zero in the first-passage protocol). Thus, qualitatively, both the first passage and finite-time protocols behave similarly in this limit.}


To summarize, when comparing various resetting protocols, we find that instantaneous distributed resetting and finite-time resetting with relaxation to a steady state give the same rate of work upon a rescaling of time. For first-passage resetting, agreement with the current finite-time method is only obtained in the limit of very sharp resetting traps. We note that while Eq. \eqref{mean_V} predicts a smaller rate of work for the finite-time scheme than the first-passage scheme, this is generally not the case; Fig.~\ref{fig:njp-curr-manu} shows the opposite case for some values of the parameter $\theta$. For brevity, Table~\ref{tab:comp} summarizes the mean rates of work for the different potentials and  resetting protocols compared in this section.

\begin{table*}
    \centering
    \begin{tabular}{|c|c|c|}\hline
        $\:$ & Harmonic potential $(V = \frac{1}{2} \lambda_V x^2)$ &  V-shaped potential $(V =\gamma_V |x|)$\\ \hline
       Finite-time  & $\cfrac{D \lambda_V}{1+r \theta \tau_\text{rel}} $ & $\dfrac{1}{1+r\theta \tau_{\rm rel}} \dfrac{4 D \sqrt{r}\alpha_V}{\sqrt{r} + 2 \sqrt{\alpha_V}}$\\\hline
        IDR \cite{mori2023entropy} & $ D \lambda_V$ & $ \dfrac{4 D \sqrt{r}\alpha_V}{\sqrt{r} + 2 \sqrt{\alpha_V}}$\\\hline
        FPR \cite{Deepak2022_work} & See~Fig.~\ref{fig:njp-curr-manu} & $\dfrac{4 D \sqrt{r}\alpha_V}{\sqrt{r} + 2 \sqrt{\alpha_V}}$ \\ \hline
    \end{tabular}
    \caption{Summary of mean rate of work for a Brownian particle with diffusivity $D$ for different resetting protocols; the current finite-time method proposed in this paper, the instantaneous distributed resetting (IDR)~\cite{mori2023entropy}, and the first-passage resetting (FPR)~\cite{Deepak2022_work}. Here, the parameter $\alpha_V\equiv \gamma_V^2/(4D)$.}
    \label{tab:comp}
\end{table*}

\section{Conclusion}\label{sec:concl}
In summary, this paper presents a general framework for investigating the thermodynamic cost associated with a finite-time stochastic resetting process. Under the assumption that the particle has time to relax to a Boltzmann state during each resetting phase, we calculate the moment generating function for the work, valid for any spatial dimension, any duration of the exploration phase, and any potentials used in the resetting and exploration phases. We note that while the majority of the paper considers Brownian particles relaxing to a Boltzmann state at the end of each resetting phase, the generating function we derive holds also for systems that relax to other non-equilibrium steady states. The framework presented here is therefore very general, and may be of use in a wide range of future studies. As an example, we used our framework to derive exact expressions for both the mean rate of work and the variance for switching between two harmonic potentials. We show that a minimum of the mean work can be obtained by the appropriate tuning of the stiffness of the resetting trap.

There are several possible interesting extensions of our work. For instance, in Ref.~\cite{mori2023entropy} it was shown that instantaneous stochastic resetting in a harmonic potential leads to a condensation transition in the large deviation regime of the entropy production. It would be interesting to investigate whether similar non-equilibrium transitions can be observed in the case of finite-time resetting protocols.  
{\color{black}Additionally, in stochastic resetting systems implemented by switching potentials, (stationary and transient) work-fluctuation theorems and its symmetry properties are interesting avenues for future investigation. Further, since Eq.~\eqref{trap-cks} is a general expression for the moment generating function, it would be interesting to see how different resetting time distributions (such as gamma, exponential, and periodic) influence the fluctuations of work. }

Moreover, we have shown in section \ref{sec:finite} that the mean rate of work can be calculated without the assumption that the system relaxes to a steady state. It would be useful to extend this to higher moments or indeed the full distribution to better understand the associated fluctuations in the general case of arbitrary fluctuating potentials. {\color{black}It would also be interesting to extend the framework here to the case where it takes a finite time to switch between the two potentials, as was considered in  recent experiments involving information erasure~\cite{goerlich2023experimental} and swift equilibration protocol~\cite{goerlich2024resetting}}. Finally, our work establishes an analytical framework to describe relevant experimental implementations of stochastic resetting. Therefore, it would be relevant to verify our results in experiments, for instance using colloidal particles within time-dependent laser traps.

\section{Acknowledgments}
K.S.O  and D.G acknowledge the Nordita fellowship program. Nordita is partially supported by Nordforsk. KSO acknowledges support by the Deutsche Forschungsgemeinschaft (DFG) within the project LO 418/29-1.  This work was supported by a Leverhulme Trust International Professorship Grant No. LIP-2020-014. S.K acknowledges the support of the Swedish Research  Council  through the grant 2021-05070.  The authors acknowledge the Nordita workshop “Fluctuations and First-Passage Problems” (April 2023), where this collaboration started. This research was enabled in part by support provided by BC DRI Group and the Digital Research Alliance of Canada (\url{www.alliancecan.ca}).

\appendix

\begin{widetext}

\section{Calculations for harmonic traps}
\label{sec:misc}
For the potentials considered in Sec.~\ref{opt-trap}, there are various contributions needed to derive the first two cumulants of the work  [Eqs.~\eqref{sc-mean} and \eqref{sc-var}] using Eqs.~\eqref{mean} and \eqref{sec-mom}. {Mainly, there are four contributions: 1) $\langle \Delta \phi(x)\rangle_{\rm eq}$, 2) $\langle [\Delta \phi(x)]^2\rangle_{\rm eq}$, 3) $\langle \Delta \phi(y)\rangle(t)$, and 4) $\langle [\Delta \phi(x)-\Delta  \phi(y)]^{2}\big\rangle(t)$, where $\Delta \phi(\cdot)\equiv U(\cdot) - V(\cdot)$ is the difference of the potential energies. Here, the subscript ``eq'' indicates the equilibrium average with respect to the stiff-potential (resetting potential):
\begin{align}
    \langle [\cdots] \rangle_{\rm eq} \equiv \int_{-\infty}^{+\infty}~dx~p_{\rm eq}(x) [\cdots]=\dfrac{1}{Z_V}\int_{-\infty}^{+\infty}~dx~e^{-\beta V(x)} [\cdots]\ , \label{eq-avg}
\end{align}
for the partition function $Z_V\equiv \int_{-\infty}^{+\infty}~dx~e^{-\beta V(x)}$. The time-dependent averages $\langle [\cdots](x,y) \rangle(t)$ are performed over those trajectories which started from equilibrium distribution with respect to the resetting potential $V(x)$, i.e., $p_{\rm eq}(x)$, and then, propagated in the shallow potential $U(x)$ for the time-interval $t$ [see also Eq.~\eqref{neq-avg-main}]:
\begin{align}
   \langle [\cdots](x,y) \rangle(t)\equiv  \int_{-\infty}^{+\infty}~dx~p_{\rm eq}(x) \int_{-\infty}^{+\infty}~dy~p(y,t|x)~\bigg([\cdots](x,y)\bigg)\ ,  \label{neq-avg}
\end{align}
where $p(y,t|x)$ is the probability density function of particle in the shallow potential $U(\cdot)$ to be at $y$ in time-interval $t$ starting from $x$, where the latter is distributed according to $p_{\rm eq}(x)$.

For the harmonic exploration and resetting potentials, we respectively have
\begin{align}
    U(x) &= \dfrac{1}{2}\lambda_U x^2 \label{pot-1}\\
    V(x) &= \dfrac{1}{2}\lambda_V (x-z)^2\ ,  \label{pot-2}
\end{align}
where $z$ is the resetting location.

Using the information given in Eqs.~\eqref{eq-avg}, \eqref{neq-avg}, \eqref{pot-1}, and \eqref{pot-2}, we can easily show the following expressions that help to compute the first two work's cumulants:
} 
\begin{align}
    \langle \Delta \phi(x)\rangle_{\rm eq} &= \dfrac{\gamma D (\lambda_U-\lambda_V)+z^2\lambda_U\lambda_V}{2\lambda_V}\ , \label{phi-eq}\\
    \langle [\Delta \phi(x)]^2\rangle_{\rm eq}&=\frac{3 \gamma ^2 D^2 (\lambda_U-\lambda_V)^2+z^4\lambda_U^2 \lambda_V^2 +2 \gamma D z^2 \lambda_U \lambda_V (3 \lambda_U-\lambda_V)}{4 \lambda_V^2}\ , \label{phi2-eq}\\
    \mathcal{L}_{t\to (s+r)}[\langle \Delta \phi(y)\rangle(t)] &= \dfrac{1}{2 \lambda_V (r+s) [\lambda_U+\gamma  (r+s)] [2 \lambda_U+\gamma  (r+s)]}\nonumber\\
    &\times\bigg[\gamma  D(\lambda_U-\lambda_V) [\lambda_U+\gamma  (r+s)] [2 \lambda_V+\gamma  (r+s) \nonumber\\
    &+\lambda_U \lambda_V z^2 \big\{\gamma  (r+s) [\lambda_U+\gamma  (r+s)]-2 \lambda_U \lambda_V\big\}\bigg]\ , \label{phi-neq}\\
   \mathcal{L}_{t\to (s+r)}\big[\big\langle [\Delta \phi(x)-\Delta & \phi(y)]^{2}\big\rangle(t)\big]  = \dfrac{2(\gamma D z^2 T_1 + T_2)}{T_3}\ , \label{phi2-neq}
\end{align}
where we introduced the shorthand notations:
\begin{align}
T_1 &= \lambda_V^2 \big[-6 \lambda_U^2+7 \gamma ^2 (r+s)^2+13 \gamma  \lambda_U (r+s)\big]+\lambda_V \big[-6 \lambda_U^3+\gamma ^3 (r+s)^3-4 \gamma ^2 \lambda_U (r+s)^2\nonumber\\
&-20 \gamma  \lambda_U^2 (r+s)\big]+6 \lambda_U^2 [\lambda_U+\gamma  (r+s)] [3 \lambda_U+\gamma  (r+s)]+9 \lambda_V^3 [2 \lambda_U+\gamma  (r+s)]\ ,\\
T_2 &= \gamma ^2 D^2 (\lambda_U-\lambda_V)^2 [\lambda_U+\gamma  (r+s)] [3 \lambda_U+\gamma  (r+s)] \big[3 \lambda_U^2-2 \lambda_U \lambda_V+\lambda_V (3 \lambda_V+\gamma  (r+s))\big]\nonumber\\
&+\lambda_U^4 \lambda_V^2 z^4 \big[3 (\lambda_U+\lambda_V)^2+\gamma ^2 r^2+\gamma  r (4 \lambda_U+3 \lambda_V+2 \gamma  s)+\gamma ^2 s^2+\gamma  s (4 \lambda_U+3 \lambda_V)\big]\ , \\
T_3 &= \lambda_V^2 (r+s) [\lambda_U+\gamma  (r+s)] [2 \lambda_U+\gamma  (r+s)] [3 \lambda_U+\gamma  (r+s)] [4 \lambda_U+\gamma  (r+s)] \ .
\end{align}

{Substituting $\langle \Delta \phi(x)\rangle_{\rm eq}$~\eqref{phi-eq} and $\mathcal{L}_{t\to (s+r)}[\langle \Delta \phi(y)\rangle(t)]$~\eqref{phi-neq} in Eq.~\eqref{mean} gives the Laplace transform of the mean work. Similarly, using Eqs.~\eqref{phi-eq}--\eqref{phi2-neq} and the Laplace transform of the mean work in Eq.~\eqref{sec-mom}, we get the Laplace transform of the second moment of the work.}

\section{Terms contributing to Eq.~\eqref{sc-var}}
\label{sec:var-exp}
{In the previous Appendix~\ref{sec:misc}, we have discussed the computation of the Laplace transformed first two moments of the work. Using these moments, we can compute the scaled variance of the work, given in Eq.~\eqref{sc-var} in the main text, and it has the following contributions:} 
\begin{align}
\Sigma_1 &\equiv  \frac{r}{\lambda_V^2 (\lambda_U+\gamma  r)^3 (2 \lambda_U+\gamma  r)^3 (3 \lambda_U+\gamma  r) (4 \lambda_U+\gamma  r) (\theta  r \tau_{\rm rel} +1)^3}\ , \\
\Sigma_2 &\equiv \gamma ^2 D^2 (\lambda_U-\lambda_V)^2 (\lambda_U+\gamma  r)^3 (3 \lambda_U+\gamma  r) \big(8 \lambda_U^2 \big(3 \lambda_U^2-2 \lambda_U \lambda_V+3 \lambda_V^2\big)\nonumber\\
&+2 \gamma ^3 \theta ^2 \lambda_V r^5 \tau_{\rm rel} ^2+\gamma ^2 \theta  r^4 \tau_{\rm rel}  \big(4 \gamma  \lambda_V+\theta  \tau_{\rm rel}  \big(5 \lambda_U^2+6 \lambda_U \lambda_V+5 \lambda_V^2\big)\big)+2 \gamma  r^3 \big(\gamma ^2 \lambda_V\nonumber\\
&+4 \gamma  \theta  \tau_{\rm rel}  (\lambda_U+\lambda_V)^2+\theta ^2 \lambda_U \tau_{\rm rel} ^2 \big(9 \lambda_U^2+2 \lambda_U \lambda_V+9 \lambda_V^2\big)\big)+4 r^2 \big(\gamma ^2 (\lambda_U+\lambda_V)^2\nonumber\\
&+\gamma  \theta  \lambda_U \tau_{\rm rel}  \big(7 \lambda_U^2+6 \lambda_U \lambda_V+7 \lambda_V^2\big)+4 \theta ^2 \lambda_U^2 \tau_{\rm rel} ^2 \big(\lambda_U^2+\lambda_V^2\big)\big)\nonumber\\
&+8 \lambda_U r \big(\gamma  \big(2 \lambda_U^2+\lambda_U \lambda_V+2 \lambda_V^2\big)+4 \theta  \lambda_U \tau_{\rm rel}  \big(\lambda_U^2+\lambda_V^2\big)\big)\big) \ ,\\
\Sigma_3 &\equiv \lambda_U^4 \lambda_V^2 z^4 \big(24 \lambda_U^4 (\lambda_U+\lambda_V)^2+\gamma ^6 \theta ^2 r^8 \tau_{\rm rel} ^2+4 \gamma ^5 \theta ^2 r^7 \tau_{\rm rel} ^2 (2 \lambda_U+\lambda_V)\nonumber\\
&+\gamma ^4 \theta  r^6 \tau_{\rm rel}  \big(\gamma  (2 \lambda_V-4 \lambda_U)+\theta  \tau_{\rm rel}  \big(24 \lambda_U^2+26 \lambda_U \lambda_V+5 \lambda_V^2\big)\big)\nonumber\\
&+2 \gamma ^3 \theta  r^5 \tau_{\rm rel}  \big(\gamma  \big(-12 \lambda_U^2-2 \lambda_U \lambda_V+3 \lambda_V^2\big)+\theta  \lambda_U \tau_{\rm rel}  \big(17 \lambda_U^2+30 \lambda_U \lambda_V+13 \lambda_V^2\big)\big)\nonumber\\
&+\gamma ^2 r^4 \big(2 \gamma ^2 (\lambda_V-2 \lambda_U)^2-2 \gamma  \theta  \lambda_U \tau_{\rm rel}  \big(24 \lambda_U^2+29 \lambda_U \lambda_V\nonumber\\
&-9 \lambda_V^2\big)+\theta ^2 \lambda_U^2 \tau_{\rm rel} ^2 \big(23 \lambda_U^2+58 \lambda_U \lambda_V+43 \lambda_V^2\big)\big)+2 \gamma  \lambda_U r^3 \big(\gamma ^2 \big(24 \lambda_U^2-14 \lambda_U \lambda_V+\lambda_V^2\big)\nonumber\\
&-2 \gamma  \theta  \lambda_U \tau_{\rm rel}  \big(10 \lambda_U^2+27 \lambda_U \lambda_V+\lambda_V^2\big)+\theta ^2 \lambda_U^2 \tau_{\rm rel} ^2 \big(3 \lambda_U^2+10 \lambda_U \lambda_V+11 \lambda_V^2\big)\big)\nonumber\\
&+4 \gamma  \lambda_U^2 r^2 \big(\gamma  \big(24 \lambda_U^2+\lambda_U \lambda_V-3 \lambda_V^2\big)-\theta  \lambda_U \tau_{\rm rel}  \big(3 \lambda_U^2+14 \lambda_U \lambda_V+7 \lambda_V^2\big)\big)\nonumber\\
&+8 \gamma  \lambda_U^4 r (10 \lambda_U+9 \lambda_V)\big)\ ,\\
\Sigma_4 &\equiv 2 \gamma  D \lambda_U^2 \lambda_V z^2 (\lambda_U+\gamma  r) \big(24 \lambda_U^4 \big(3 \lambda_U^3-\lambda_U^2 \lambda_V-\lambda_U \lambda_V^2+3 \lambda_V^3\big)+\gamma ^6 \theta ^2 \lambda_V r^8 \tau_{\rm rel} ^2\nonumber\\
&+\gamma ^5 \theta  r^7 \tau_{\rm rel}  \big(2 \gamma  \lambda_V+\theta  \tau_{\rm rel}  \big(5 \lambda_U^2+3 \lambda_U \lambda_V+6 \lambda_V^2\big)\big)+\gamma ^4 r^6 \big(\gamma ^2 \lambda_V+2 \gamma  \theta  \tau_{\rm rel}  \big(4 \lambda_U^2+5 \lambda_U \lambda_V+5 \lambda_V^2\big)\nonumber\\
&+\theta ^2 \tau_{\rm rel} ^2 \big(43 \lambda_U^3-11 \lambda_U^2 \lambda_V+39 \lambda_U \lambda_V^2+8 \lambda_V^3\big)\big)+\gamma ^3 r^5 \big(\gamma ^2 \big(4 \lambda_U^2+5 \lambda_U \lambda_V+5 \lambda_V^2\big)\nonumber\\
&+\gamma  \theta  \tau_{\rm rel}  \big(68 \lambda_U^3+11 \lambda_U^2 \lambda_V+66 \lambda_U \lambda_V^2+13 \lambda_V^3\big)+\theta ^2 \lambda_U \tau_{\rm rel} ^2 \big(141 \lambda_U^3-54 \lambda_U^2 \lambda_V+90 \lambda_U \lambda_V^2+53 \lambda_V^3\big)\big)\nonumber\\
&+\gamma ^2 r^4 \big(\gamma ^2 \big(36 \lambda_U^3+\lambda_U^2 \lambda_V+36 \lambda_U \lambda_V^2+6 \lambda_V^3\big)+\gamma  \theta  \lambda_U \tau_{\rm rel}  \big(228 \lambda_U^3-25 \lambda_U^2 \lambda_V\nonumber\\
&+176 \lambda_U \lambda_V^2+81 \lambda_V^3\big)+\theta ^2 \lambda_U^2 \tau_{\rm rel} ^2 \big(221 \lambda_U^3-71 \lambda_U^2 \lambda_V+87 \lambda_U \lambda_V^2+127 \lambda_V^3\big)\big)\nonumber\\
&+\gamma  \lambda_U r^3 \big(\gamma ^2 \big(132 \lambda_U^3-51 \lambda_U^2 \lambda_V+111 \lambda_U \lambda_V^2+38 \lambda_V^3\big)+2 \gamma  \theta  \lambda_U \tau_{\rm rel}  \big(190 \lambda_U^3-41 \lambda_U^2 \lambda_V+120 \lambda_U \lambda_V^2\nonumber\\
&+95 \lambda_V^3\big)+2 \theta ^2 \lambda_U^2 \tau_{\rm rel} ^2 \big(83 \lambda_U^3-15 \lambda_U^2 \lambda_V+15 \lambda_U \lambda_V^2+65 \lambda_V^3\big)\big)+2 \lambda_U^2 r^2 \big(\gamma ^2 \big(122 \lambda_U^3-73 \lambda_U^2 \lambda_V\nonumber\\
&+84 \lambda_U \lambda_V^2+49 \lambda_V^3\big)+2 \gamma  \theta  \lambda_U \tau_{\rm rel}  \big(77 \lambda_U^3-15 \lambda_U^2 \lambda_V+33 \lambda_U \lambda_V^2+53 \lambda_V^3\big)+24 \theta ^2 \lambda_U^2 \tau_{\rm rel} ^2 \big(\lambda_U^3+\lambda_V^3\big)\big)\nonumber\\
&+4 \lambda_U^3 r \big(\gamma  \big(54 \lambda_U^3-32 \lambda_U^2 \lambda_V+19 \lambda_U \lambda_V^2+33 \lambda_V^3\big)+24 \theta  \lambda_U \tau_{\rm rel}  \big(\lambda_U^3+\lambda_V^3\big)\big)\big)\ .
\end{align}

\end{widetext}

\section{Derivation of Eq. (\ref{eq:work_V})}\label{sec:Vcalc}
Let us consider the case when the particle is freely diffusing in the exploration phase, and is reset using a V-shaped potential $V(x) = \gamma_V |x|$. Then, we have
\begin{align}
    &\Delta \phi(x) \equiv -\gamma_V |x|\ ,\\
    &p_{\rm eq}(x) = \dfrac{\beta \gamma_V}{2}e^{-\beta \gamma_V|x|}\ .
\end{align}

The equilibrium mean of $\Delta \phi(x)$ can be obtained as
\begin{align}
    \langle \Delta \phi(x)\rangle_{\rm eq} = -1/\beta\ .  \label{phieq}
\end{align}

Next we compute the non-equilibrium average of $\Delta \phi(x)$ as
\begin{align}
    \langle \Delta \phi(y)\rangle(t) = \int_{-\infty}^{+\infty}~\rmd x~p_{\rm eq}(x)~\underbrace{\int_{-\infty}^{+\infty}~\rmd y~p(y,t|x)~\Delta \phi(y)}_{Q(x)}\ . \label{qteqn}
\end{align}
To proceed further, we define for convenience the integral
\begin{align}
    Q(x) &\equiv \int_{-\infty}^{+\infty}~\rmd y~p(y,t|x)~\Delta \phi(y)\ ,\\
    &=-\dfrac{\gamma_V}{\sqrt{4\pi D t}}\int_{0}^{+\infty}~dy~y\bigg[e^{-(y + x)^2/(4 D t)} + e^{-(y-x)^2/(4 D t)}\bigg]\ .
\end{align}

Defining shorthand notations $a_1 \equiv \gamma_V/\sqrt{4 \pi D t}$, $a_2 \equiv 4 D t$, so that
\begin{align}
    Q(x) &=-a_1\int_{0}^{+\infty}~dy~y\bigg[e^{-(y + x)^2/a_2} + e^{-(y-x)^2/a_2}\bigg]\ ,\\
    & = -a_1\bigg(\sqrt{\pi a_2}~x~\text{Erf}\left(\frac{x}{\sqrt{a_2}}\right)+a_2 e^{-\frac{x^2}{a_2}}\bigg)\ , \label{q2eqn}
\end{align}
where $\text{Erf}(u)\equiv \frac{2}{\sqrt{\pi}}\int_0^u~\rmd t~e^{-t^2} $ is the error function.

We then substitute~$Q(x)$~\eqref{q2eqn} in Eq.~\eqref{qteqn} and perform the integral over $x$ by noticing that $Q(x)$ is an even function of $x$. This gives
\begin{align}
    \langle \Delta \phi(y)\rangle(t) &= \int_{-\infty}^{+\infty}~dx~p_{\rm eq}(x)~Q(x)\ ,\\
    &= -a_1 \dfrac{\sqrt{\pi } \sqrt{a_2} e^{\frac{a_2 a_3^2}{4}} \text{Erfc}\left(\frac{\sqrt{a_2} a_3}{2}\right)+a_2 a_3}{a_3}\ \label{phineq},
\end{align}
for $a_3 \equiv \beta \gamma_V$.

Substituting the values of $\langle\Delta \phi\rangle_{\rm eq}$~\eqref{phieq} and $\langle\Delta \phi\rangle(t)$~\eqref{phineq} in Eq.~(\ref{mean}) in the main text  gives the scaled mean work, as given in Eq.~\eqref{eq:work_V}.



%

\end{document}